\documentclass[conference,compsoc]{IEEEtran}

\ifCLASSOPTIONcompsoc
  \usepackage[nocompress]{cite}
\else
  \usepackage{cite}
\fi

\usepackage{nicefrac}
\usepackage{siunitx}
\usepackage{array,framed}
\usepackage{booktabs}
\usepackage{tabularx}
\usepackage{
  color,
  float,
  epsfig,
  wrapfig,
  graphics,
  graphicx,
  subcaption
}

\usepackage{textcomp,amssymb}
\usepackage{setspace}
\usepackage{latexsym,fancyhdr,url}
\usepackage{enumerate}
\usepackage{enumitem}
\usepackage[most]{tcolorbox}

\usepackage{algorithm}
\usepackage[noend]{algpseudocode}
\usepackage[symbol]{footmisc}

\usepackage[numbers,sort]{natbib}
\usepackage{tikz}

\newcommand*\circled[1]{\tikz[baseline=(char.base)]{
            \node[shape=circle,draw=blue, fill=blue,inner sep=1pt] (char) {\textcolor{white}{\bfseries#1}};}}

\usepackage{graphics}
\usepackage{xparse} 
\usepackage{xspace}
\usepackage{multirow}
\usepackage{csvsimple}
\usepackage{balance}
\usepackage{pifont}
\usepackage{svg}

\usepackage{
  tikz,
  pgfplots,
  pgfplotstable
}
\usepackage{xcolor}
\definecolor{darkgreen}{rgb}{0.0, 0.5, 0.0} 
\usepackage[hidelinks]{hyperref}
 \hypersetup{
     colorlinks=true,
     citecolor=darkgreen,     
     linkcolor=red!70!black,       
     urlcolor=blue!70!black       
 }
\usepackage[capitalise,nameinlink, noabbrev]{cleveref}

\newcommand{\email}[1]{{\rm\textsf{\href{mailto:#1}{#1}}}}

\definecolor{darkviolet}{HTML}{9400D3}

\newcommand{\failure}[2]{\circled{#1} {\textcolor{blue}{{#2}}}\xspace}

\newcommand{\failanalog}{\failure{3}{Analog Malfunction}}

\newcommand{\failvrms}{\failure{1}{VRM Malfunctions}}
\newcommand{\faildigitals}{\failure{2}{Digital Malfunctions}}
\newcommand{\failanalogs}{\failure{3}{Analog Malfunctions}}
\newcommand{\failcaps}{\failure{4}{Capacitor Malfunctions}}
\usetikzlibrary{
  shapes.geometric,
  arrows,
  external,
  pgfplots.groupplots,
  matrix
}

\pgfplotsset{compat=1.9}

\usepackage{mathtools}

\DeclareMathAlphabet{\mathcal}{OMS}{cmsy}{m}{n}

\usepackage{svg}

\DeclareGraphicsExtensions{%
    .png,.PNG,%
    .pdf,.PDF,%
    .jpg,.mps,.jpeg,.jbig2,.jb2,.JPG,.JPEG,.JBIG2,.JB2}

\usepackage{xparse}
\newcommand{\bnm}{\begin{newmath}}
\newcommand{\enm}{\end{newmath}}

\newcommand{\bea}{\begin{eqnarray*}}%
\newcommand{\eea}{\end{eqnarray*}}%

\newcommand{\bne}{\begin{newequation}}
\newcommand{\ene}{\end{newequation}}

\newcommand{\bal}{\begin{newalign}}
\newcommand{\eal}{\end{newalign}}

\newenvironment{newalign}{\begin{align}%
\setlength{\abovedisplayskip}{4pt}%
\setlength{\belowdisplayskip}{4pt}%
\setlength{\abovedisplayshortskip}{6pt}%
\setlength{\belowdisplayshortskip}{6pt} }{\end{align}}

\newenvironment{newmath}{\begin{displaymath}%
\setlength{\abovedisplayskip}{4pt}%
\setlength{\belowdisplayskip}{4pt}%
\setlength{\abovedisplayshortskip}{6pt}%
\setlength{\belowdisplayshortskip}{6pt} }{\end{displaymath}}

\newenvironment{newequation}{\begin{equation}%
\setlength{\abovedisplayskip}{4pt}%
\setlength{\belowdisplayskip}{4pt}%
\setlength{\abovedisplayshortskip}{6pt}%
\setlength{\belowdisplayshortskip}{6pt} }{\end{equation}}

\newcounter{ctr}

\newcounter{mytable}
\def\mytable{\begin{centering}\refstepcounter{mytable}}
\def\endmytable{\end{centering}}

\newcounter{myfig}
\def\myfig{\begin{centering}\refstepcounter{myfig}}
\def\endmyfig{\end{centering}}

\newlength{\saveparindent}
\newlength{\saveparskip}
\newcommand{\E}{{\rm I\kern-.3em E}}

\renewcommand{\eqref}[1]{\mbox{Equation~(\ref{#1})}}

\newcommand{\mV}{m\kern-1pt V\xspace}
\newcommand{\Vpp}{V\kern-1pt \textsubscript{pp}\xspace}

\def \part {part}

\renewcommand{\paragraph}[1]{\vspace*{6pt}\noindent\textbf{#1}\;}

\def \blackslug{\hbox{\hskip 1pt \vrule width 4pt height 8pt
    depth 1.5pt \hskip 1pt}}
\def \qed{\quad\blackslug\lower 8.5pt\null\par}

\newcounter{mynote}[section]

\newcommand\ignore[1]{}

\newcounter{rcnote}[section]

\newcounter{mrnote}[section]

\newcounter{fknote}[section]

\newcounter{anote}[section]

\DeclareMathSymbol{\mlq}{\mathord}{operators}{``}
\DeclareMathSymbol{\mrq}{\mathord}{operators}{`'}

\newcommand{\rhf}[2]{R_{f, \gamma}}

\DeclareDocumentCommand{\edist}{o o}{
  \ensuremath{
    \IfNoValueTF{#1}{{d}}{{\sf d}(#1,#2)}
  }
}

\newcommand{\olrk}[1]{\ifx\nursymbol#1\else\!\!\mskip4.5mu plus 0.5mu\left(\mskip0.5mu plus0.5mu #1\mskip1.5mu plus0.5mu \right)\fi}

\NewDocumentCommand{\indseq}{ O{1} O{r} }{{#1}\ldots {#2}}

\makeatletter
\newcommand{\linebreakand}{%
  \end{@IEEEauthorhalign}
  \hfill\mbox{}\par
  \mbox{}\hfill\begin{@IEEEauthorhalign}
}
\makeatother

\definecolor{byzantium}{rgb}{0.44, 0.16, 0.39}

\newif\ifshowcomments
\showcommentstrue 

\begin{document}
\fancyhead{}
\def\thetitle{Chypothermia: Clock Freezing for Static Side-channel Attacks} 



\title{\thetitle}


\author{\IEEEauthorblockN{Fatemeh Khojasteh Dana}
\IEEEauthorblockA{Worcester Polytechnic Institute\\
\small\email{fdana@wpi.edu}}
\and
\IEEEauthorblockN{Mehmet Ali Cetin}
\IEEEauthorblockA{Worcester Polytechnic Institute\\
\small\email{mcetin@wpi.edu}}
\and
\IEEEauthorblockN{Xinrui Wang}
\IEEEauthorblockA{Ruhr University Bochum\\
\small\email{xinrui.wang@rub.de}}
\linebreakand
\IEEEauthorblockN{Andrew Butler}
\IEEEauthorblockA{Worcester Polytechnic Institute\\
\small\email{abutler@wpi.edu}}
\and
\IEEEauthorblockN{Yuval Yarom}
\IEEEauthorblockA{Ruhr University Bochum\\
\small\email{yuval.yarom@rub.de}}
\and
\IEEEauthorblockN{Shahin Tajik}
\IEEEauthorblockA{Worcester Polytechnic Institute\\
\small\email{stajik@wpi.edu}}}

\date{}

\maketitle

\thispagestyle{plain} 
\pagestyle{plain}

\begin{abstract}

Static side-channel attacks, which exploit halted-clock conditions to extract sensitive information, pose an increasing threat to chip security. 
To counter these attacks, various defenses have been proposed that monitor for abnormal clock behavior and trigger the clearing of sensitive data when clock anomalies are detected.
In this work, we demonstrate that exposing a chip to cryogenic temperatures interferes with on-chip mixed-signal components, responsible for signal sensing and generation. 
Based on this observation, we develop \emph{Chypothermia}, an attack that, without any electrical tampering with the system, disables the target clock sensor, the clock generation circuit, and the voltage sensors, while preserving the secret data.

While effective at halting the clock, cooling is a slow process and, on its own, is often insufficient against systems equipped with temperature sensors designed to detect thermal anomalies. 
To bypass these protections, we combine Chypothermia with Chypnosis (Mitard et al., IEEE SP 2026) and show that, even within a moderately low-temperature operating range, this combination can halt the clock while evading detection. 
We implement Chypothermia on multiple FPGA/SoC platforms and demonstrate successful disabling of both soft-IP and hard-IP sensor implementations. 
Finally, we apply Chypothermia to the alert handler of the OpenTitan root of trust, which incorporates a state-of-the-art clock sensor, and show that the attack evades detection and prevents key zeroization.
Finally, we introduce and implement an FPGA-compatible self-heating sensor as a countermeasure and demonstrate its robustness against Chypothermia.

\end{abstract}

\section{Introduction}\label{sec:intro}
Over the last three decades, various physical side-channels have been discovered, showing that the security of cryptographic implementations on chips can be undermined.
The primary assumption for most physical side-channels is that information leakage occurs during data transitions.
Deploying data randomization or shuffling in countermeasures, such as masking and hiding~\cite{koblah2022hardware}, is a conventional technique to mitigate dynamic side-channel attacks,
as it prevents repetition and integration of measurements
over multiple clock cycles. 
However, there has been a rise in a class of attacks called static side-channel attacks, in which attackers halt the chip's clock and recover static data stored in memory components, such as Flip-Flops (FFs). 
In this case, randomness becomes ineffective if the adversary halts the circuit to probe it between two clock cycles using attacks such as static power analysis~\cite{moradi2014side}, Laser Logic State Imaging (LLSI)~\cite{krachenfels2021real}, Impedance Analysis (IA)~\cite{monfared2023leakyohm}, and Thermal Laser Stimulation (TLS)~\cite{krachenfels2021automatic}.

Because recovering static data stored in registers takes significantly longer than a clock cycle, stopping the circuit clock is the main requirement for static attacks~\cite{krachenfels2021real,monfared2023leakyohm}.
Consequently, countermeasures that detect clock tampering~\cite{dumitru2025borrowed,farheen2022twofold} and respond by clearing sensitive state are generally effective against such attacks.
Furthermore, direct manipulation of the system clock is often impractical in real-world settings because many secure integrated circuits rely on internal clock sources for cryptographic operations~\cite{microchip_security}. 
As a result, attacks that can halt the system without requiring direct access to the clock are particularly relevant.

\begin{figure}[t]
    \centering
    \includegraphics[width=\linewidth]{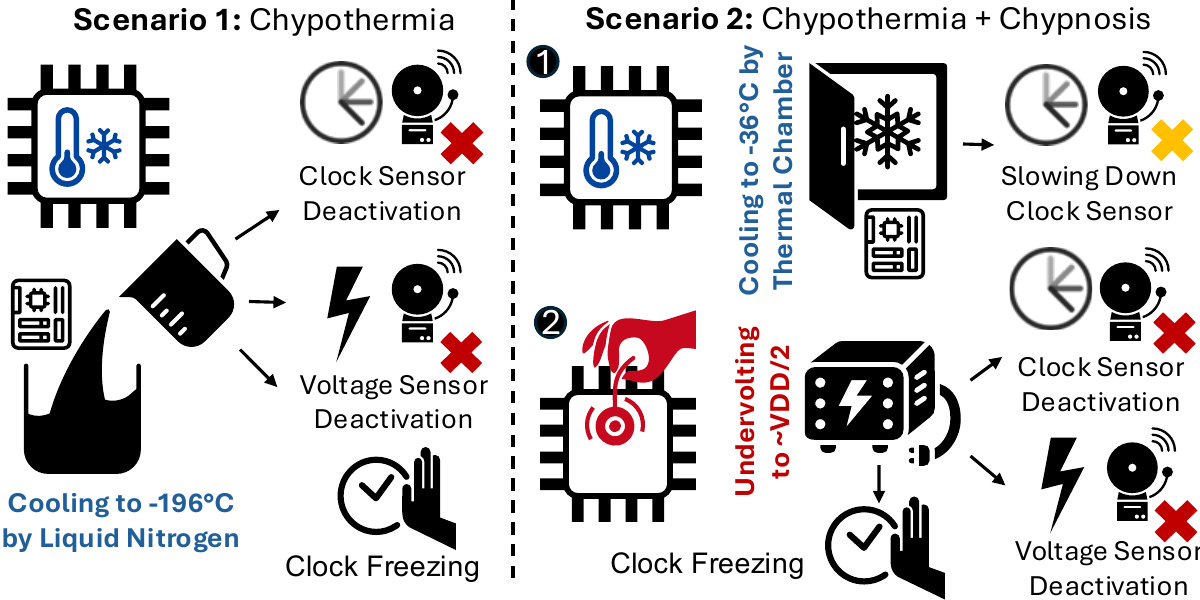}
    \caption{High-level overview of Chypothermia attack scenarios}
    \label{fig:chypothermia}
\end{figure}

Recently, Chypnosis~\cite{mitard2025chypnosis} demonstrated that rapidly dropping the supply voltage to nearly half of its nominal value can place a chip into a hibernation state, in which clock propagation ceases while sensitive data remains intact and can subsequently be recovered using static attacks. 
The work showed that although the analog sensing component of existing clock sensors detects a clock stoppage, the digital response logic cannot clear sensitive data quickly enough before the clock halts. 
To address this limitation, follow-up countermeasures based on asynchronous clearing were proposed to respond more rapidly to undervolting attacks~\cite{mitard2025chypnosis,AMD-SB-8018,PSIRT-118}. 
However, it remains an open question whether an adversary can non-invasively manipulate the analog sensing component itself without using invasive techniques~\cite{helfmeier2013breaking, yamashita2022redshift}.

On-chip analog components, such as sensors and oscillators, are usually designed to operate robustly in various environmental conditions.
However, if the adversary operates the chip in extremely cold or hot temperatures, these analog components might start operating erratically or completely fail to function.
Operating the chip in a hot environment might not be feasible, as it could cause a meltdown and irreversible damage.
However, operating the chip in extremely cold temperatures has been shown to be safe.
The digital fabrics of devices, such as Field Programmable Gate Arrays (FPGAs), have been reported to operate reliably at cryogenic temperatures (below -150$^{\circ}$C) across various applications, including deep-space missions~\cite{sheldon2011cryogenic,Bakhsi2011}, quantum computing~\cite{homulle2016cryogenic,conway2016fpga,morris2023fingerprinting}, and superconducting magnets for particle acceleration~\cite {Turqueti2020cryogenic}.
However, it has also been reported that both on-chip mixed-signal modules (e.g., internal clock generation circuits and sensors) and off-chip components (e.g., decoupling capacitors and voltage regulators) could become unreliable~\cite{sheldon2011cryogenic,homulle2018design}.

Although most anti-tamper mechanisms monitor voltage and clock anomalies, they are typically not configured to respond to temperature fluctuations~\cite{microchip_security}, particularly low-temperature conditions. 
One reason is that the same device may be deployed across a wide range of operating environments, e.g., an FPGA embedded in a drone operating at different altitudes and in different geographic locations, making it difficult to distinguish legitimate environmental variations from deliberate temperature manipulation. 
Furthermore, even when temperature sensors are configured to detect excessively low temperatures, the threshold is constrained by the intended deployment environment. 
For example, for applications operating in cold regions, such as the Arctic~\cite{jaffe2026arctic}, the threshold cannot be set above certain temperatures (e.g., -40$^{\circ}$C) without risking false alarms and zeroization during normal operation.

Driven by these observations, we ask the following questions:
\emph{(1) Is it possible to halt the system’s
clock at cryogenic temperatures without electrically tampering with its source? (2) Can we halt the clock at moderate cold temperatures without triggering state-of-the-art clock sensors?}

\noindent\textbf{Our Contribution.}
In this work, we answer both questions affirmatively. We
introduce \textit{Chypothermia} attacks, in which an adversary puts the chip in a cryogenic condition, bypasses the clock sensors, and deactivates the on-chip clock sources while the digital fabric of the chip retains the data. 
The adversary can then warm the chip while keeping the clock halted to perform a static side-channel attack (e.g., LLSI) and recover the retained data. 
Our attack exploits the observation that at a cryogenic temperature (i.e., -196$^{\circ}$C)  the on-chip mixed-signal components, such as Phase-Locked Loops (PLLs) and analog-to-digital converters (ADCs), fail, while on-chip digital circuits continue to function and retain data. 
\cref{fig:chypothermia} presents an abstract overview of our attack.

Although cryogenic temperatures can stop the clock, cooling the chip is a slow process; hence, systems equipped with temperature sensors, intended to detect temperature anomalies, can detect the attack.
To evade detection, we also present a combined attack, in which we mount Chypothermia at a moderately cold temperature within the temperature sensor's allowed thresholds (above -40$^{\circ}$C) to reduce the sensor's sensitivity and then mount Chypnosis~\cite{mitard2025chypnosis} to put the chip in hibernation without triggering the clock sensor.

We perform Chypothermia on various SRAM- and Flash-based FPGAs/SoCs fabricated in 28\,nm and 16\,nm processes.
First, we conduct extensive experiments using liquid nitrogen (LN$_{2}$) to determine the time required to induce failure in PLLs and ADCs at cryogenic temperatures. 
Next, we demonstrate that staying in cryogenic conditions effectively halts the clock without requiring direct control of the clock source or triggering the clock sensor.
To demonstrate the effectiveness of Chypothermia in practical scenarios, we integrate the state-of-the-art sensor proposed in~\cite{mitard2025chypnosis} into the FPGA implementation of OpenTitan English Breakfast~\cite{opentitan_englishbreakfast} and successfully stop the clock of its side-channel-protected cryptographic engine without triggering zeroization.
Finally, we propose and implement a circuit-based, FPGA-compatible self-heating countermeasure that maintains the sensor's temperature within the nominal range and mitigates the attack.


\begin{figure*}[t]
    \centering
    \includegraphics[width=0.85\linewidth]{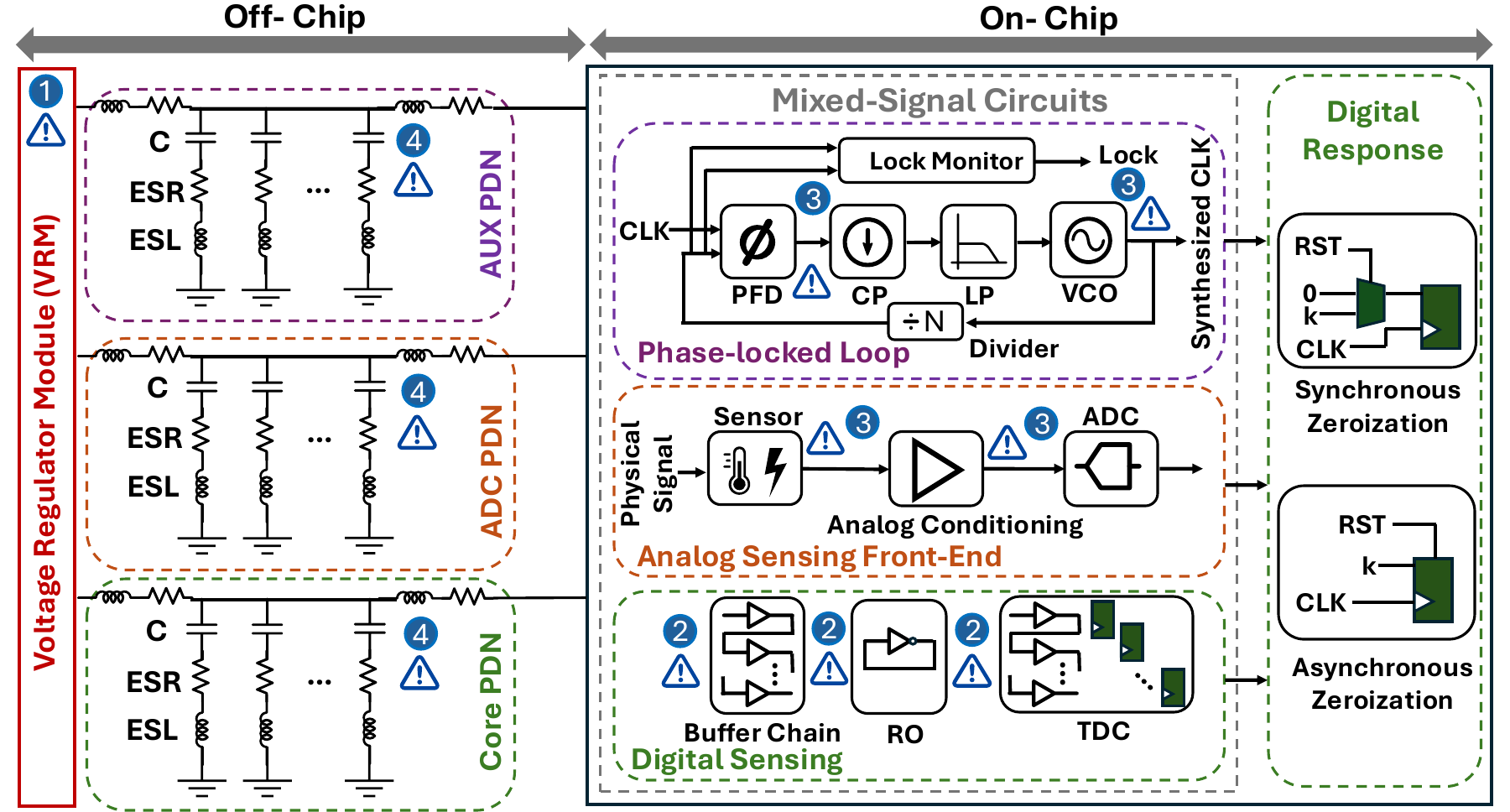}
    \caption{On modern FPGAs, PLLs, sensors, and digital cores have their own PDNs. Possible failures at extremely cold temperatures, indicated by blue warning signs, include: \textbf{1. VRM Malfunction}, \textbf{2. Digital Malfunction}, \textbf{3. Analog Malfunction}, and \textbf{4. Capacitor Malfunction}.}
    \label{fig:pdn}
\end{figure*}

\section{System Operation in Cryogenic Conditions}\label{sec:chip_in_cryogenic}

\subsection{Digital Circuits}
Chip vendors specify operating ranges for temperature and voltage in their public datasheets. 
However, these specified ranges are usually very conservative, and most digital circuits still function beyond them.
While there is a limit to how hot a chip can get, digital chips have been shown to function reliably at extremely low temperatures, such as -267$^{\circ}$C (4K)~\cite{sheldon2011cryogenic,Bakhsi2011,homulle2016cryogenic,conway2016fpga,Turqueti2020cryogenic}.

Two temperature-dependent competing factors in CMOS logic offset each other, enabling digital circuits to remain stable and function under cryogenic conditions. 
Reducing temperature increases electron mobility, thereby increasing the electrical current~\cite{sheldon2011cryogenic}.
Conversely, cold temperatures increase the threshold voltage for both PMOS and NMOS transistors, reducing gate overdrive and current and making it harder to turn on a transistor~\cite{sheldon2011cryogenic,Turqueti2020cryogenic}.
However, depending on the transistor technology, one of these two factors is slightly more dominant than the other.
In older, larger transistor technologies, cooling the circuit makes electron mobility the dominant factor; hence, transistors draw more current and switch faster.
In this case, voltage regulator modules (VRMs) might not be able to keep up with the chip's current surge, leading to significant voltage drops in the core voltage domain. 
We refer failures caused by such voltage drops as \failvrms, see \cref{fig:pdn}.

However, in modern technology nodes (e.g., sub 28\,nm), we observe inverse temperature dependence (ITD)~\cite{glamovcanin2023temperature,neshatpour2018enhancing}, in which the circuit slows at lower temperatures.
ITD could have security implications for FPGAs, as some soft IP FPGA sensors exploit the propagation delay of digital logic to detect anomalies. 
Buffer chain-based~\cite{BorrowedTime_github}, ring oscillators (ROs), and Time-to-digital converters (TDCs) are examples of such delay-based sensors.
As a result, longer delays in these sensors at cold temperatures relative to their calibrated values could reduce their reaction time. 
We refer these failures as \faildigitals, see \cref{fig:pdn}.

\subsection{Mixed-Signal Circuits}
Mixed-signal circuits, such as PLLs and ADCs, combine analog and digital components to process or synthesize real-world signals.
In contrast to digital blocks on a chip, mixed-signal circuits are more customized; therefore, their response to extremely cold temperatures varies.
However, generally, cryogenic conditions reshape the analog circuit's operating point due to changes in threshold voltage, mobility, transconductance, leakage current, and parasitic impedances.
The result is that bias, reference, startup, and timing margins move first, and eventually the macro-block fails as the supporting circuits drift out of range.
We refer these failures as \failanalogs, see \cref{fig:pdn}.

\subsection{Printed Circuit Boards (PCBs)}
The Power Distribution Network (PDN) on a PCB is responsible for reliably delivering voltage to the chip. 
Commercial components on the PDN include discrete components (e.g., decoupling capacitors) and integrated circuits (e.g., DC-DC voltage regulators), all of which are designed for specific temperature ranges.
Their failure at cryogenic temperatures could cause voltage and current fluctuations, leading to destabilized chip behavior.
Commercial Surface Mount Device (SMD) capacitors can experience a significant 90\% drop in capacitance at cryogenic temperatures, and their parasitic resistance (ESR) can increase up to 1000$\times$~\cite{teyssandier2010commercially, homulle2017reconfigurable}.
In this case, mixed-signal modules on chips suffer more as their PDN is usually separate from the core voltage PDN, and, therefore, the  available  capacitance of the PDN under normal conditions is much lower and, thus, more sensitive to  capacitance loss~\cite{homulle2019cryogenic}.
We call failures caused by such capacitance loss as \failcaps, see \cref{fig:pdn}.
Certain voltage regulators have also been shown to misfunction below -196$^{\circ}$C (77K)~\cite{homulle2018design}. These are, again, \failvrms.

\section{Clock Sources and Sensors}
\subsection{Clock Synthesis}
The system clock can be generated from various on-chip or off-chip sources.
The most common sources are off-chip crystal, as well as internal RC and ring oscillators~\cite{farheen2022twofold}.
Regardless of the clock source, these sources can usually generate only a limited number of fixed frequencies, which is insufficient for the clock requirements of various IPs on the chip.
Moreover, some of these sources are suffering from jitter and are sensitive to environmental conditions.

To synthesize multiple stable clock frequencies and reduce clock skew, on-chip PLLs are deployed.
The main building blocks of PLLs are the phase frequency detector (PFD), charge pump, loop filter, and voltage-controlled oscillator (VCO), see \cref{fig:pdn}.
While PFDs are digital, Charge Pumps and VCOs are analog, and the loop filter is composed of passive components.
VCOs are tuned to operate in the device operating temperature range.
They may fail to operate outside their specified range or continue to oscillate, but there is no guarantee of frequency accuracy.

\subsection{Clock Sensors}

To detect whether a clock has stopped or become unstable, different monitoring techniques can be used.
A clock sensor consists of two main modules, namely, an analog sensing module and a digital response module.

\subsubsection{PLL-based Sensing}
One common approach is to use PLL-based sensors~\cite{dumitru2025borrowed}. A PLL has an output signal called Lock, which indicates whether the PLL is synchronized with its reference clock, see \cref{fig:pdn}. The PLL continuously compares the reference clock and the generated output clock. When the input clock stops or changes significantly, the PLL loses synchronization, and the Lock signal is deasserted. Therefore, the locked signal can be used as a simple indicator of clock failure.

\label{sec:setup}

\subsubsection{Delay-based Sensing}

Another way to detect clock-stop attacks is to use a delay-based detection circuit within the chip's digital fabric, such as a buffer-chain sensor~\cite{dumitru2025borrowed}, as shown in \cref{fig:pdn,fig:sensor}. In this approach, a chain of LUTs forms a delay path for monitoring the clock's behavior.
Under normal operation, the clock signal propagates through this delay path, and the outputs of different buffers in the chain will be "1" or "0" alternatively, depending on whether the rising or falling edge of the clock signal has passed through them.
However, when the clock is stopped, all buffers' outputs will eventually be equal. 
Such a condition can indicate that the clock has been halted and can be detected by an XNOR gate connected to the outputs of buffers.
In this case, the output of the XNOR gate is connected to the response logic and is used to trigger the protection mechanism.

\subsection{Clock Tamper Response}

After detecting that the clock is halted, synchronous or asynchronous responses can be generated.
The main issue with the synchronous response is that it requires the response to wait for the next active clock edge, see \cref{fig:pdn}. 
Such requirements cannot be met if the chip is undervolted rapidly, as in Chypnosis attack~\cite{mitard2025chypnosis}.
To speed up the response time, asynchronous preset/clear signals have been proposed to wipe the secret~\cite{mitard2025chypnosis}.
This is important because, during a clock-stop attack, the system clock may already be stopped or unreliable.
By using asynchronous paths, the countermeasure can react immediately after the delay-based detector is triggered, without relying on the system clock, which might already have stopped.
\section{Threat Model}
We assume that the attacker has physical access to the target device.
Moreover, the attacker can obtain snapshots of the hardware state using techniques such as LLSI~\cite{krachenfels2021real}, and then reconstruct the values stored in the registers.
In the case of LLSI, failure-analysis tools are often available for rental at a few hundred dollars per hour, avoiding the need for a full purchase.
For secret extraction, we assume a template-attack threat model, in which the attacker must identify the locations of the target registers, either through reverse engineering or by leveraging prior knowledge of the design.
During the attack phase, we assume that all detection- and response-based countermeasures are active.
We also assume that the internal workings of the security sensors may be unknown to the attacker. 
With these assumptions, we consider two attack scenarios.

\begin{figure*}[t]
  \setlength{\belowcaptionskip}{2pt}
  \centering
  \begin{subfigure}[b]{0.374\textwidth}
    \centering
    \includegraphics[width=\linewidth]{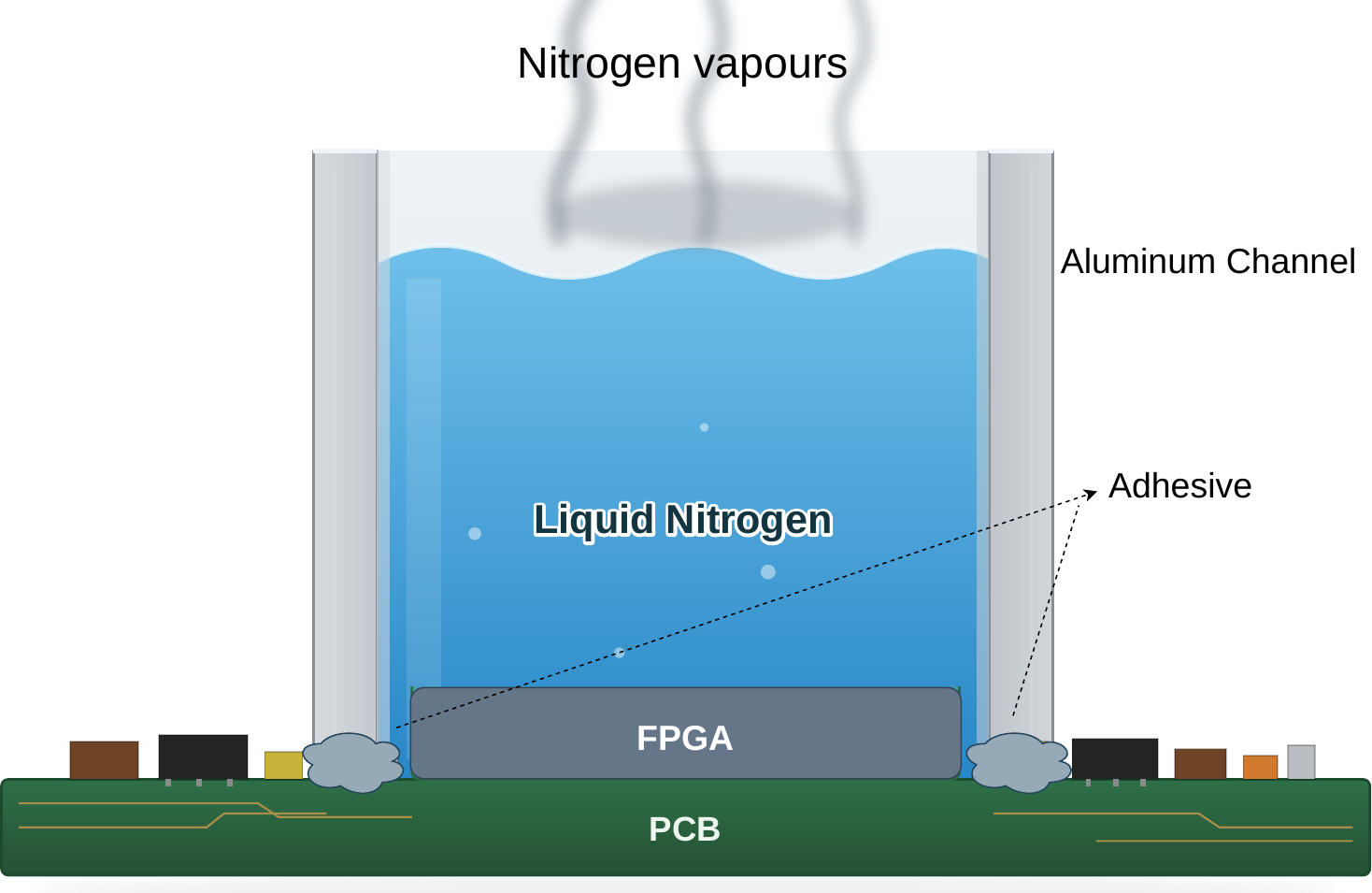}
    \caption{Cross-sectional schematic}
    \label{fig:cryo_setup}
  \end{subfigure}\hfill
  \begin{subfigure}[b]{0.333\textwidth}
    \centering
    \includegraphics[width=\linewidth]{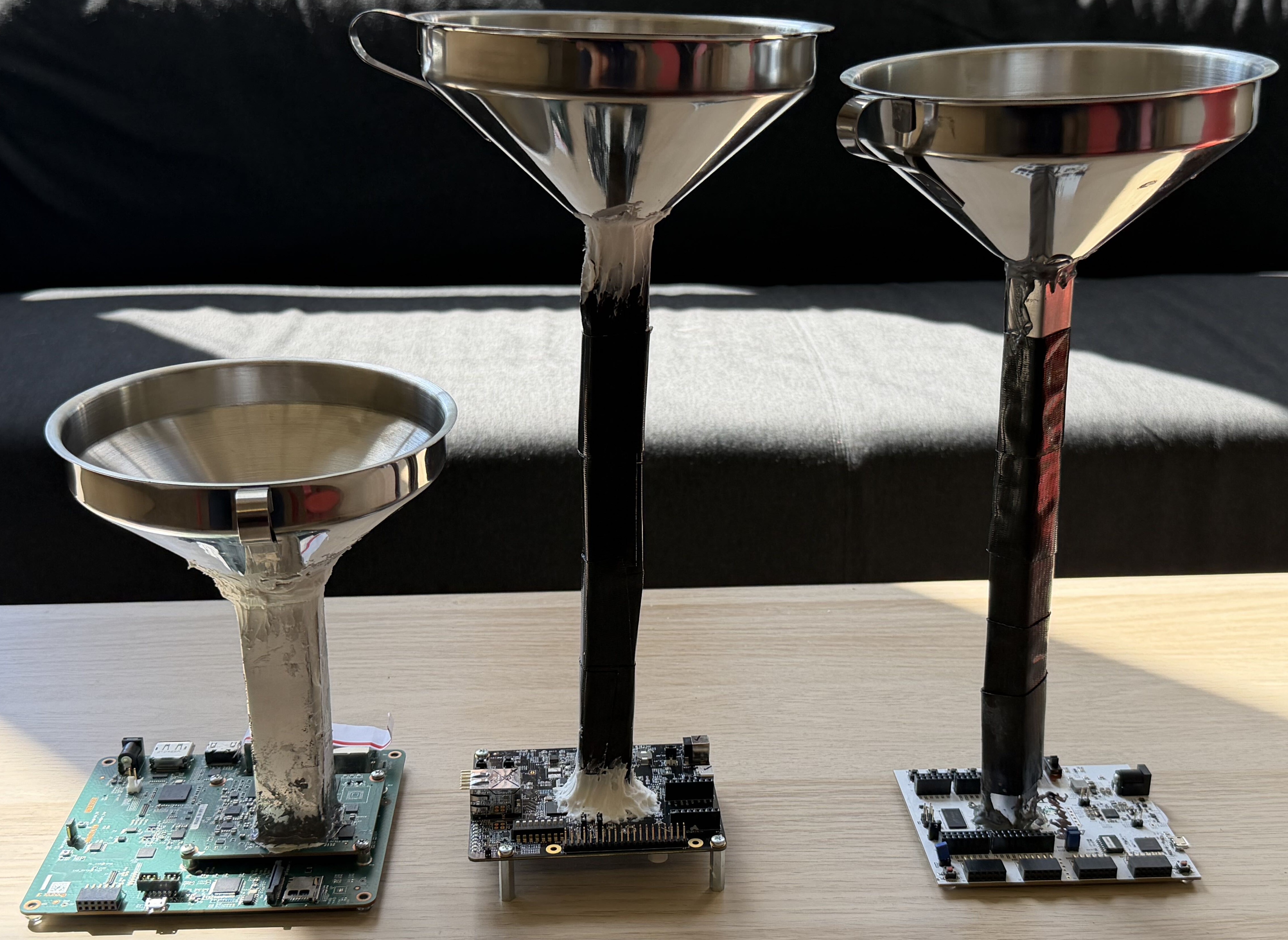}
    \caption{Fixtures on DUTs}
    \label{fig:exprimental-setup}
  \end{subfigure}\hfill
  \begin{subfigure}[b]{0.273\textwidth}
    \centering
    \includegraphics[width=\linewidth]{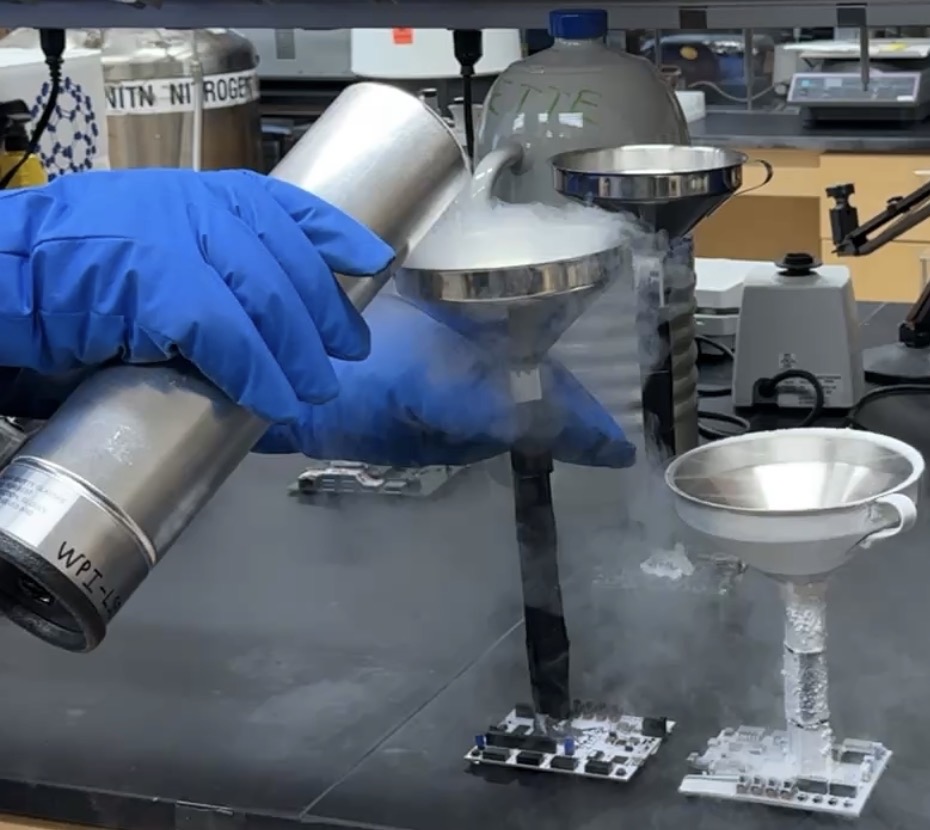}
    \caption{DUT exposure to LN$_2$}
    \label{fig:ln2-pour}
  \end{subfigure}
  \caption{Cryogenic test environment. (a) The FPGA-bearing PCB sits in an aluminum channel filled with liquid nitrogen, cooling the die to cryogenic temperatures. (b) All evaluated platforms under test, each fitted with a funnel-based LN$_2$ delivery setup. (c) LN$_2$ was poured from a dewar through a funnel onto the target board during an experiment. The target board was not wired up for an orderly demonstration.}
  \label{fig:cryo_combined}
\end{figure*}

\noindent\textbf{Scenario 1:} In the first scenario, we assume the chip is equipped with a clock sensor and that the clock is generated internally.
However, the chip either lacks a temperature sensor or its temperature sensor is not configured by the user.
This is a realistic scenario for mission-critical systems, such as FPGAs deployed on military drones, since they must operate in various conditions.
In this case, the temperature sensor is not usually deployed as an anti-tamper sensor, as its potential triggering by environmental changes can lead to system unavailability during the mission.
Such a system could fall into the adversary's hands and be exposed to cryogenic temperatures, leading to stopping the clock without triggering the clock sensor.
Liquid gases or cryocoolers can be employed to reach such cryogenic temperatures.

\noindent\textbf{Scenario 2:} In the second scenario, we assume that the clock is still generated internally, but the chip's temperature sensor has been configured as an anti-tamper sensor. 
While most temperature sensors are configured for hot temperatures to prevent fault injection or meltdown, they might not be configured for cold temperatures.
Even if they are configured for cold temperatures, the threshold cannot be set too high.
For example, the temperature in the northern regions can reach -40$^{\circ}$C; hence, for an operational system operating at such temperatures, the threshold should be set much lower than -40$^{\circ}$C to prevent false alarms.
If such a system falls into the adversary's hands, she can cool down the chip to a modest subzero temperature within the allowed ranges to slow down the clock sensor, and then undervolt (Chypnosis attack~\cite{mitard2025chypnosis}) the chip to hibernate it, and consequently, stop the clock without triggering the clock sensor.
In this scenario, the adversary can cool the chip using a thermal chamber to achieve a controlled temperature decrease. 
However, a rapid drop in temperature could also be achieved using inexpensive freeze sprays or Peltier-based thermoelectric coolers (TECs) with less control over the magnitude of the temperature drop.
We further assume that the attacker can access and modify the chip’s core voltage supply rails and remove the decoupling capacitors on the printed circuit board (PCB). Carrying out this type of tampering requires some understanding of the PCB schematic, which can be obtained from documentation, visual inspection, or multimeter measurements.

\noindent\textbf{Real-world Implications:} To see how an adversary could gain from this type of attack in practice, we can look at the following cases.
One case involves the decryption core on FPGAs or microcontrollers/microprocessors, which is configured with a cryptographic key.
These decryption cores can be employed, for instance, to decrypt the device’s bitstream, firmware, or any other confidential data.
By extracting the secret key, the adversary can clone, reverse-engineer, spoof, or tamper with the design contained in the bitstream or firmware. 
Moreover, if the same key is used across multiple chips in the field, the attacker can compromise the security of other chips that use it.

\section{Experimental Setup}

\subsection{Devices under Test (DUT)}
To test the effects of extremely cold temperatures on mixed-signal and digital circuits of chips, we used three devices from two vendors as follows. 

\subsubsection{AMD Devices}
We used AMD Kria KV260 Vision AI Starter Kits (part number SK-KV260-G)~\cite{amd_kv260}, which contain SRAM-based AMD Zynq UltraScale+ MPSoC chips.
The kit includes a K26 System-on-Module (SOM) connected to a vision carrier card and uses a fan and heatsink as its thermal solution. The AMD Zynq UltraScale+ MPSoC is packaged in a flip-chip package and soldered to the SOM, which combines an ARM processing system (PS) with programmable logic (PL) as an FPGA fabric. The device is fabricated with a 16~nm technology~\cite{amd_zynq_ultrascale_mpsoc}.
For our experiments, we removed the chip's fan and heatsink to expose the silicon backside directly to liquid nitrogen.

We also used Digilent Arty S7 boards~\cite{digilent_arty_s7_rm}, which contain AMD Spartan-7 FPGAs in a BGA package. The FPGA part number is XC7S50-CSGA324, and it is fabricated using a 28 nm technology~\cite{amd_spartan7}.
These boards were used for both liquid nitrogen and thermal chamber experiments.
Since the latter experiment was designed to test the combined attack, which involved rapid undervolting, we prepared additional samples from the same families and removed several capacitors from them to make them suitable for the attack.
Specifically, the C30, C63, C29, C28, C27, C25, C36, and C37 capacitors were removed, which means approximately 99\% of the capacitance on the core voltage rail was removed. To supply the core voltage externally, the R221 shunt resistor, which connects the voltage rail between the buck regulator and the silicon, was removed. Next, a BNC female-to-DuPont male test lead connector was soldered to the target PCB’s JP3 test pin, allowing a function generator to supply the desired core voltage externally.
For AMD Spartan-7 and Zynq UltraScale+ boards, we used the Vivado Design Suite~\cite{amd_vivado} to implement and program the FPGA designs.

\subsubsection{Microchip Devices}
We used a Microchip PolarFire Discovery Kit board, which includes a Flash-based Microchip PolarFire FPGA SoC with part number MPFS095T-1FCSG325E~\cite{microchip_PolarFire} in a BGA package. This board is suitable for embedded development, digital signal processing (DSP), and edge-computing applications. The FPGA fabric can be configured using Microchip’s Libero SoC software~\cite{microchip_libero}.

\begin{figure*}
    \centering
    \includegraphics[width=0.90\linewidth]{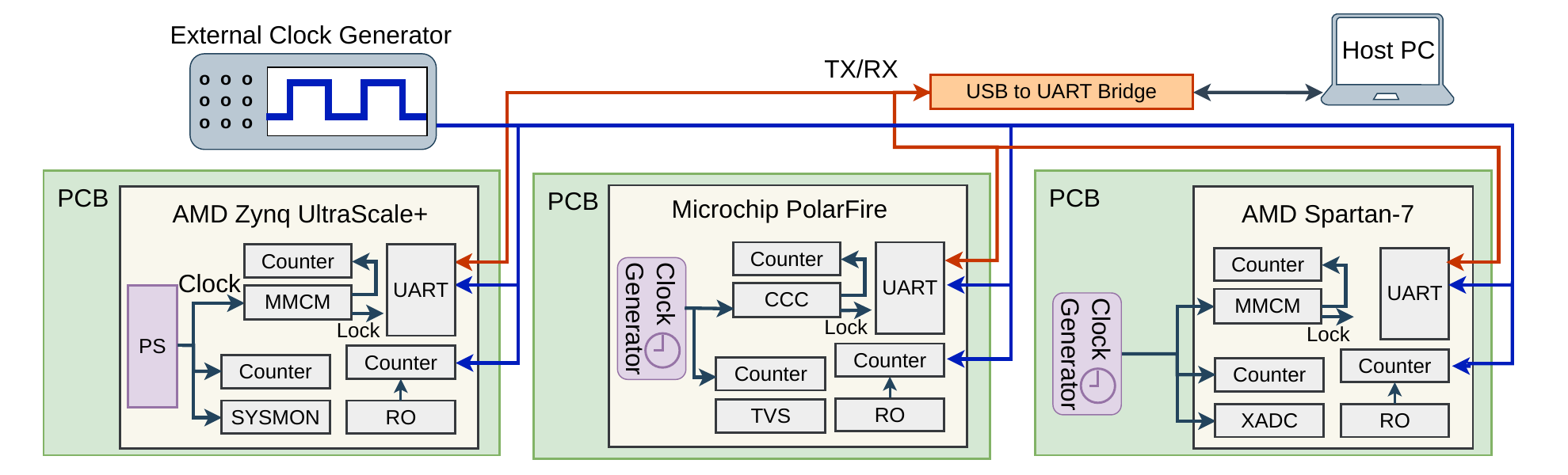}
    \caption{Experimental setup for the cryogenic evaluation. The host PC communicates with the DUT through a USB-to-UART interface (ESP-Prog) using the UART TX/RX pins, shown by the orange paths. An external clock source, a Tektronix AFG3021B function generator, provides a 12.5~MHz clock (shown by the blue paths) for communication and delay-monitoring measurements using the RO. The other clock sources are also shown in purple, including the on-board clock for AMD Spartan-7, the on-chip clock for Microchip PolarFire, and the PS-generated clock for AMD Zynq UltraScale+. 
    The DUT is placed inside the cryogenic test environment and is equipped with monitoring sensors.}
    \label{fig:cryogenic_test_environment_block_diagram}
\end{figure*}

\subsection{Cooling Setups}

\subsubsection{Cryogenic Setup}
For the cryogenic experiments, we used LN$_2$, which has a boiling temperature of approximately -196$^\circ$C.
To cool down each chip while preventing the surrounding components from being exposed to LN$_2$, we built a small open-topped channel directly on each board; see \cref{fig:cryo_setup}. 
The channel is a strip of aluminum, 0.2\,mm thick, bent by hand to enclose the device and wall off the rest of the PCB. 
We chose aluminum because, in our trials, plastic and similar materials became brittle and cracked at cryogenic temperatures, whereas the aluminum retained its mechanical properties. 
A 14\,cm (5.5\,in) stainless-steel kitchen funnel glued to the top directs the LN$_2$ onto the die.

To seal the channel to the board and keep it leak-proof through repeated cooling cycles, we combined two adhesives, Loctite Power Grab Express construction adhesive and J-B~Weld 8281 steel-reinforced epoxy, with heavy-duty duct tape applied in several overlapping layers, see \cref{fig:exprimental-setup}. 
To induce freezing, LN$_2$ is poured from a dewar onto the target device (see \cref{fig:ln2-pour}) for varying durations, allowing us to investigate the impact of cooling on chips.
In addition, because contact with LN$_2$ or the chilled surfaces can cause frostbite, special gloves should be used.

\subsubsection{Thermal Chamber Setup}
For conducting temperature-controlled experiments, we used a TestEquity Model 107 temperature chamber~\cite{testequity_107_manual}. TE-107 is a benchtop environmental chamber that provides controlled thermal conditions over a wide temperature range, from -40$^{\circ}$C to 130$^{\circ}$C. It includes digital controls for programming and monitoring the chamber temperature, which enabled us to conduct stable, repeatable experiments.

\begin{figure*}[t]
    \centering

    \begin{subfigure}[t]{0.32\textwidth}
        \centering
        \includegraphics[width=\linewidth]{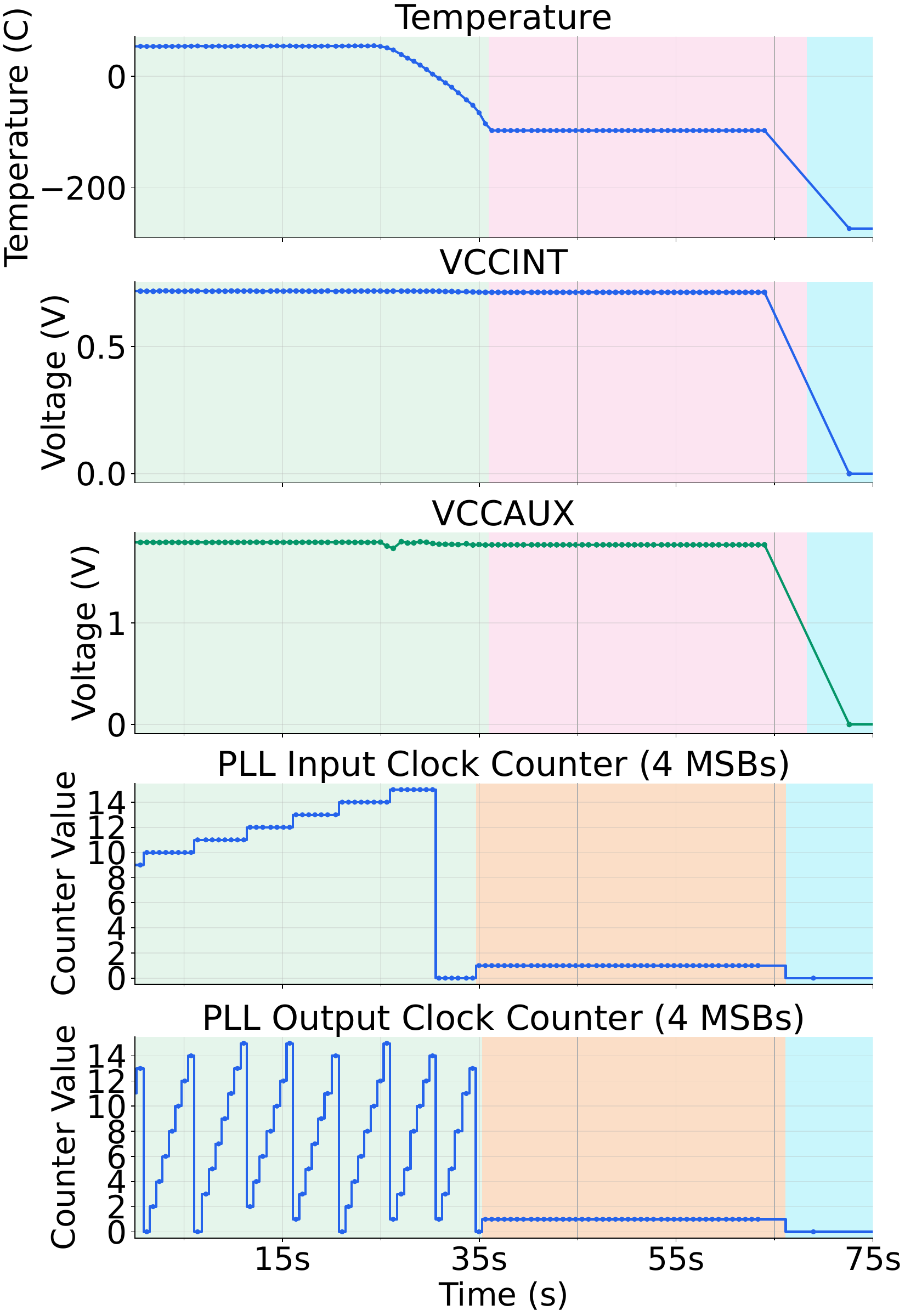}
        \caption{AMD Zynq UltraScale+ SoC}
        \label{fig:ultrascale_results}
    \end{subfigure}
    \hfill
    \begin{subfigure}[t]{0.32\textwidth}
        \centering
        \includegraphics[width=\linewidth]{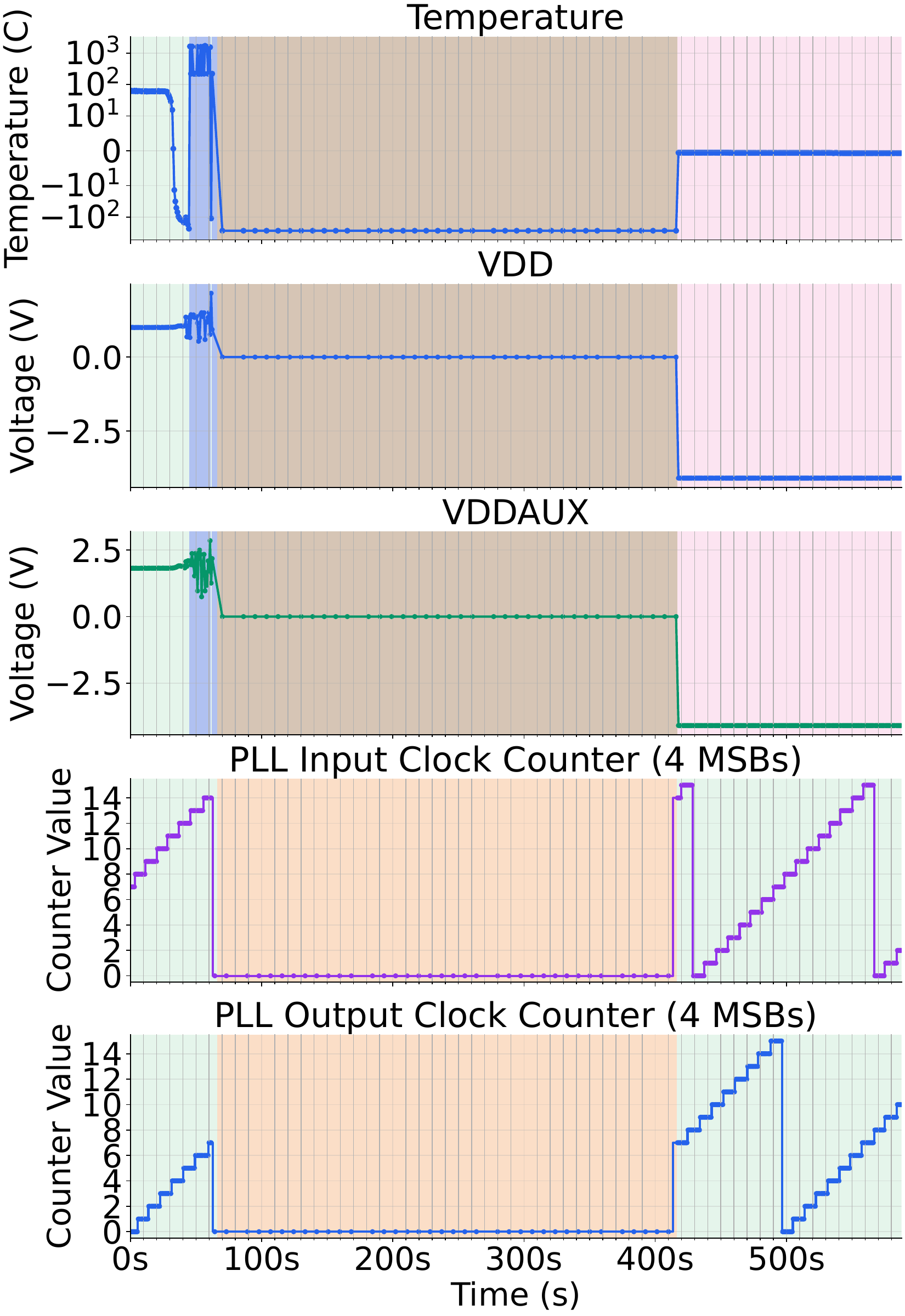}
        \caption{Microchip PolarFire SoC}
        \label{fig:PolarFire_results}
    \end{subfigure}
    \hfill
    \begin{subfigure}[t]{0.32\textwidth}
        \centering
        \includegraphics[width=\linewidth]{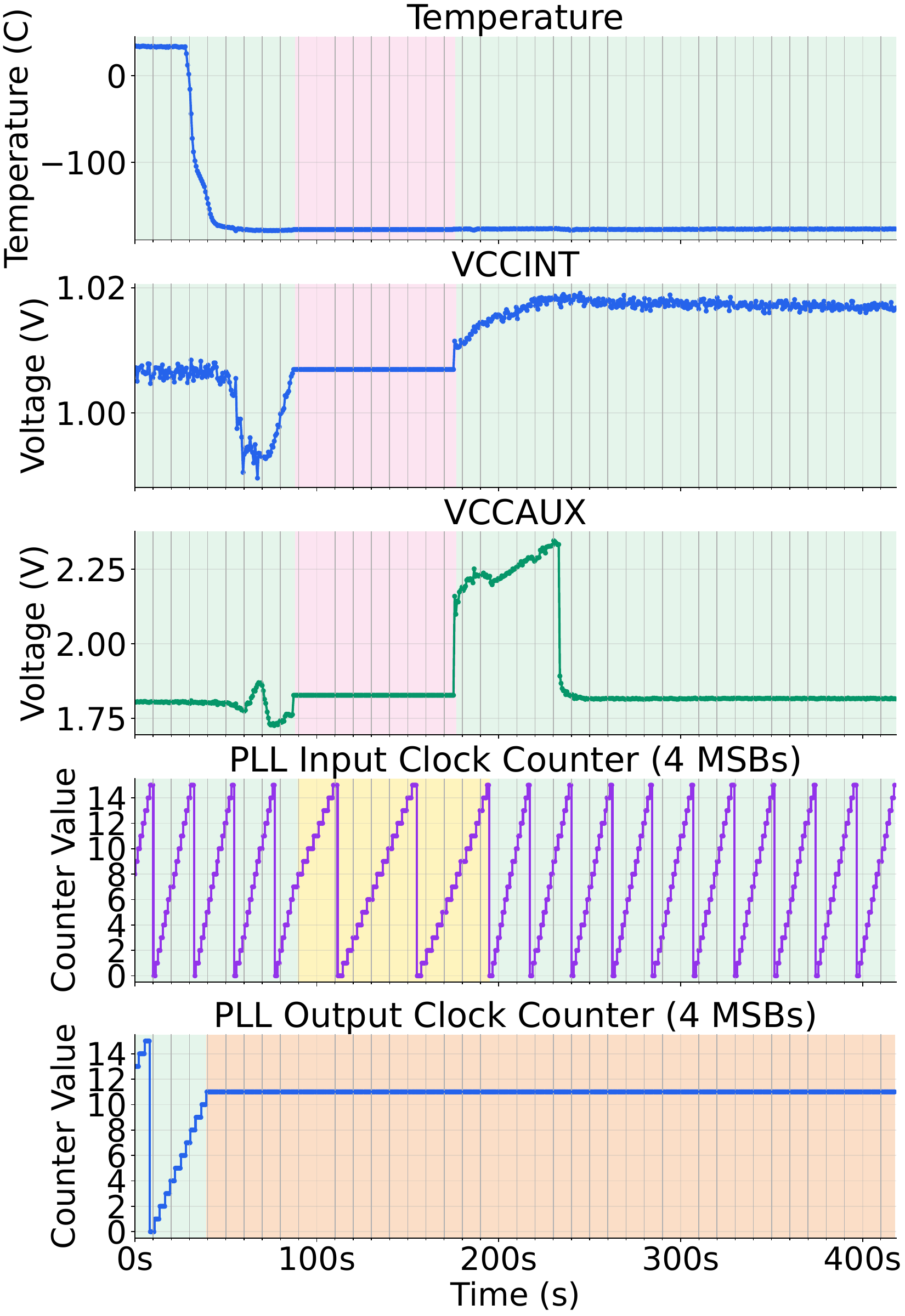}
        \caption{AMD Spartan-7 FPGA}
        \label{fig:spartan7_results}
    \end{subfigure}

    \vspace{0.9em}

    \begin{subfigure}[t]{\textwidth}
        \centering
        \includegraphics[width=\linewidth]{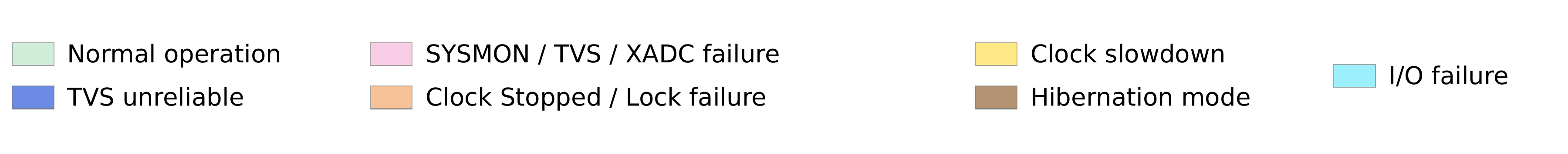}
        \label{fig:ultrascale_ro}
    \end{subfigure}
    \caption{Monitoring behavior of the DUTs under cryogenic conditions. From top to bottom, the rows show the reported temperature, monitored voltage rails, and clock counters over time for the (a) AMD Zynq UltraScale+ MPSoC, (b) Microchip PolarFire SoC, and (c) AMD Spartan-7 FPGA. The 4 most significant bits of the 28-bit counters were monitored for the MMCM/CCC clock signals to verify that the clocks remained running during the experiment.}
    \label{fig:monitoring_results_all}
\end{figure*}

\subsection{Electrical Setup}
To ensure reliable telemetry data communication for the chip under cryogenic conditions and to avoid potential failures in the on-board USB-to-serial UART interface, we used an external clock signal and FPGA I/Os as UART pins.
For our experiments, we generated an external clock from a Tektronix AFG3021B function generator~\cite{tektronix_afg3000_manual}. 
In our setup, we configured the output clock frequency to 12.5 MHz.
The UART pins were connected to an ESP-Prog board~\cite{espressif_esp_prog_guide}, which was used to communicate with the chip from the host PC (see \cref{fig:cryogenic_test_environment_block_diagram}).
For the AMD Spartan-7 and Zynq UltraScale+ boards, we configured PMOD pins to both supply the clock and communicate via UART TX/RX. For the Microchip PolarFire board, we used the Raspberry Pi header pins to both supply the clock and communicate via UART TX/RX.

In addition, we used the same Tektronix AFG3021 function generator model~\cite{tektronix_afg3000_manual} to generate the fast voltage-drop signal for our combined attack. The function generator allowed us to create the rapid voltage drop required to perform the Chypnosis attack.

\subsection{On-Chip Telemetry}
To understand the behavior of various components on our DUTs, we configured the chips to monitor the available telemetry data during the experiments.
The available data includes the on-die temperature, voltage levels on different rails, clock behavior, and digital logic delay variations. 
To monitor the temperature and voltage rails, we used the available on-chip sensing blocks. These include System Monitor (SYSMON) for the AMD UltraScale+ device, Temperature and Voltage Sensor (TVS) for the Microchip PolarFire device, and Analog-to-Digital Converter (XADC) for the AMD Spartan-7 device.

To monitor clock behavior, we implemented a 28-bit counter to count clock edges within the chip. By tracking the counter values, we were able to observe whether the clock was running normally, slowing down, or stopping under cryogenic conditions.
To monitor delay changes during our experiments, we implemented an 11-stage RO. 
The RO was built by connecting 11 Look-up tables (LUTs) configured as inverter gates in series, with the output of the last inverter gate connected back to the input of the first, forming a feedback loop.
We also added a 32-bit counter to measure the RO output frequency. The counter counted the number of edges from the RO output during a fixed observation window of 100000 clock cycles. 
The reference clock for this measurement was the external clock generated by the function generator.
The RO helped us monitor timing-delay changes in the FPGA fabric during the experiments. 
It also provided a simple way to verify that the digital fabric and I/Os were still operating correctly at cryogenic temperatures. 
If the RO continued to oscillate and the counter produced the expected values, it indicated that the chip remained functional and healthy under those conditions.

\section{Chypothermia Attack Results}\label{sec:results}

In this section, we evaluate telemetry data from the implemented sensors across three chips under cryogenic conditions. The goal is to monitor the behavior of mixed-signal and digital circuits under cryogenic attacks. 

\subsection{AMD Zynq UltraScale+ Results}

\subsubsection{AMD System Monitor Configuration}
The System Monitor (SYSMON) includes an ADC for monitoring internal device conditions, such as die temperature and various voltage rails~\cite {amd_sysmon_ug580}. 
The sampled ADC data is stored in status registers and can be accessed through different interfaces, including the Dynamic Reconfiguration Port (DRP), a JTAG interface, an I2C interface, the Power Management Bus (PMBus), and the Advanced Peripheral Bus (APB) in AMD Zynq UltraScale+ MPSoC devices.
In our design, we instantiated the System Management module in our Verilog design to monitor the die temperature, PL VCCINT, which supplies the FPGA core, and PL VCCAUX, which supplies the PLLs/MMCMs.
SYSMON has an input clock that drives its internal operation and interface logic. 

\subsubsection{Clock Configurations}
For the AMD Zynq UltraScale+ board, we used the internal clock generated by the Zynq UltraScale+ MPSoC IP. This clock was routed from the Processing System (PS) side to the Programmable Logic side, and its output frequency was configured to 33.33~MHz.
For the SYSMON and Mixed-Mode Clock Managers (MMCM) input clock, we used an internal clock generated by the PS. 

\begin{figure*}[t]
    \centering

    \begin{subfigure}[t]{0.32\textwidth}
        \centering
        \includegraphics[width=\linewidth]{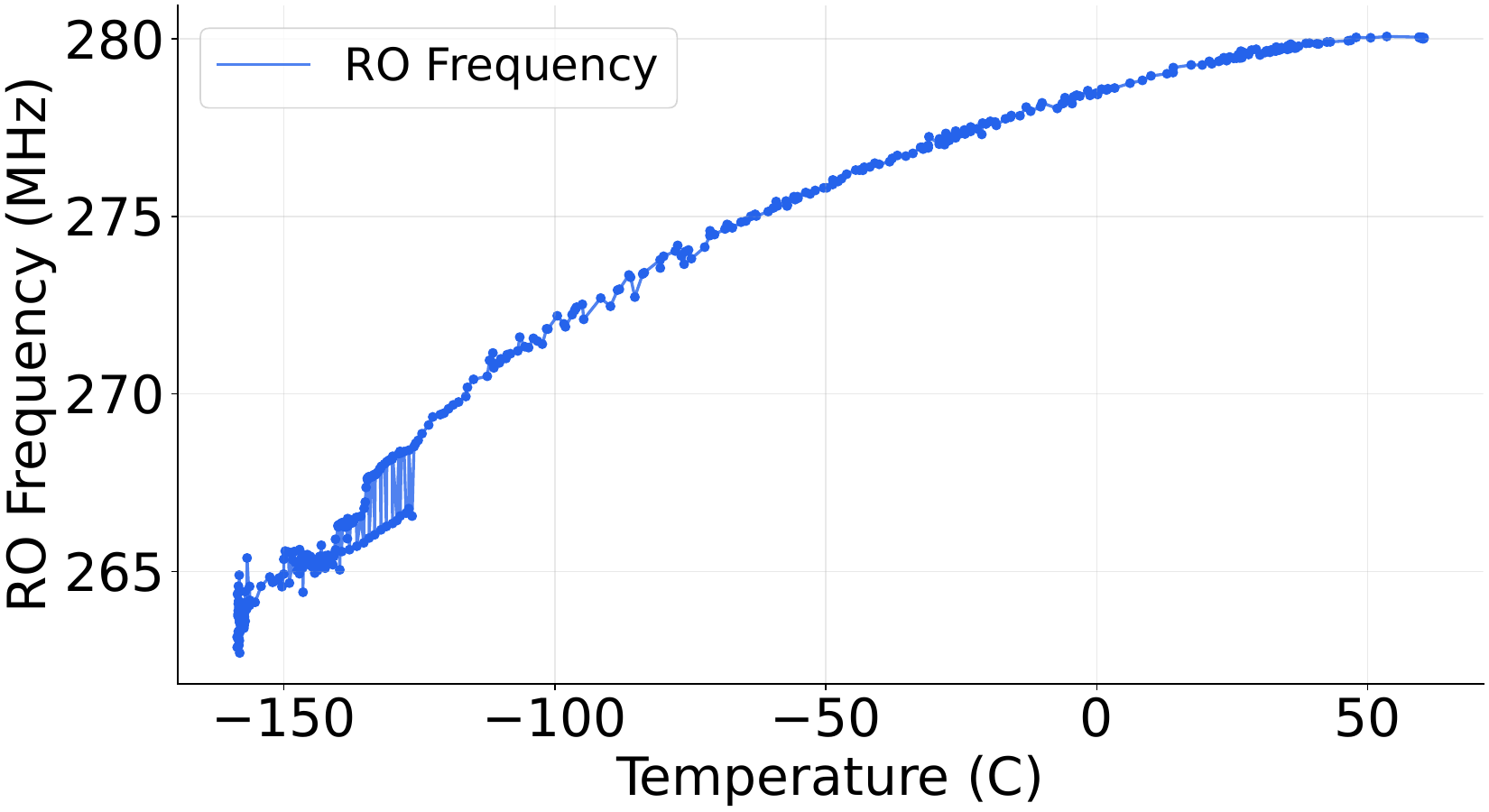}
        \caption{AMD Zynq UltraScale+}
        \label{fig:kv260_ro_temp}
    \end{subfigure}
    \hfill
    \begin{subfigure}[t]{0.32\textwidth}
        \centering
        \includegraphics[width=\linewidth]{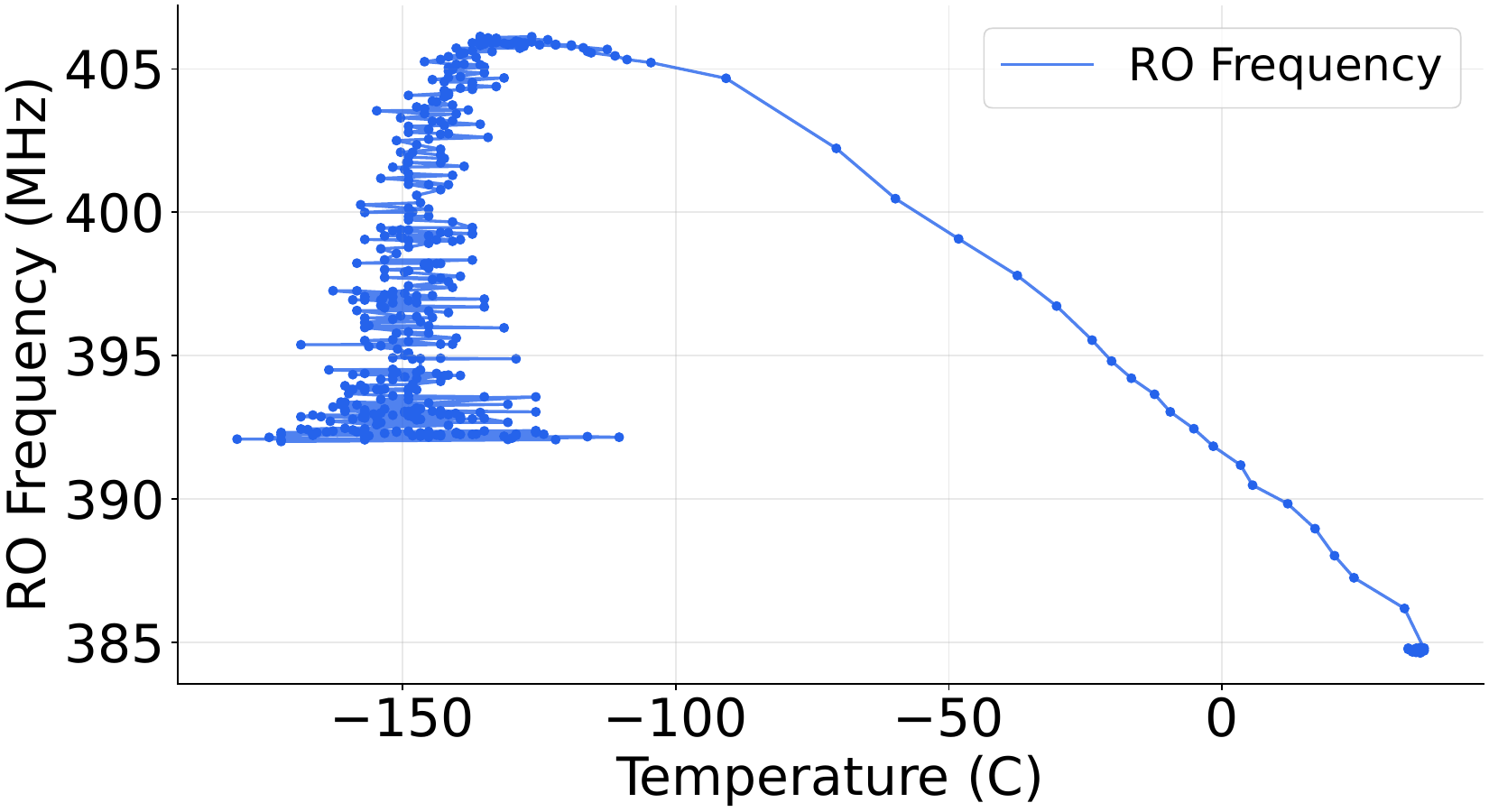}
        \caption{Microchip PolarFire}
        \label{fig:pf_ro_temp}
    \end{subfigure}
     \begin{subfigure}[t]{0.32\textwidth}
        \centering
        \includegraphics[width=\linewidth]{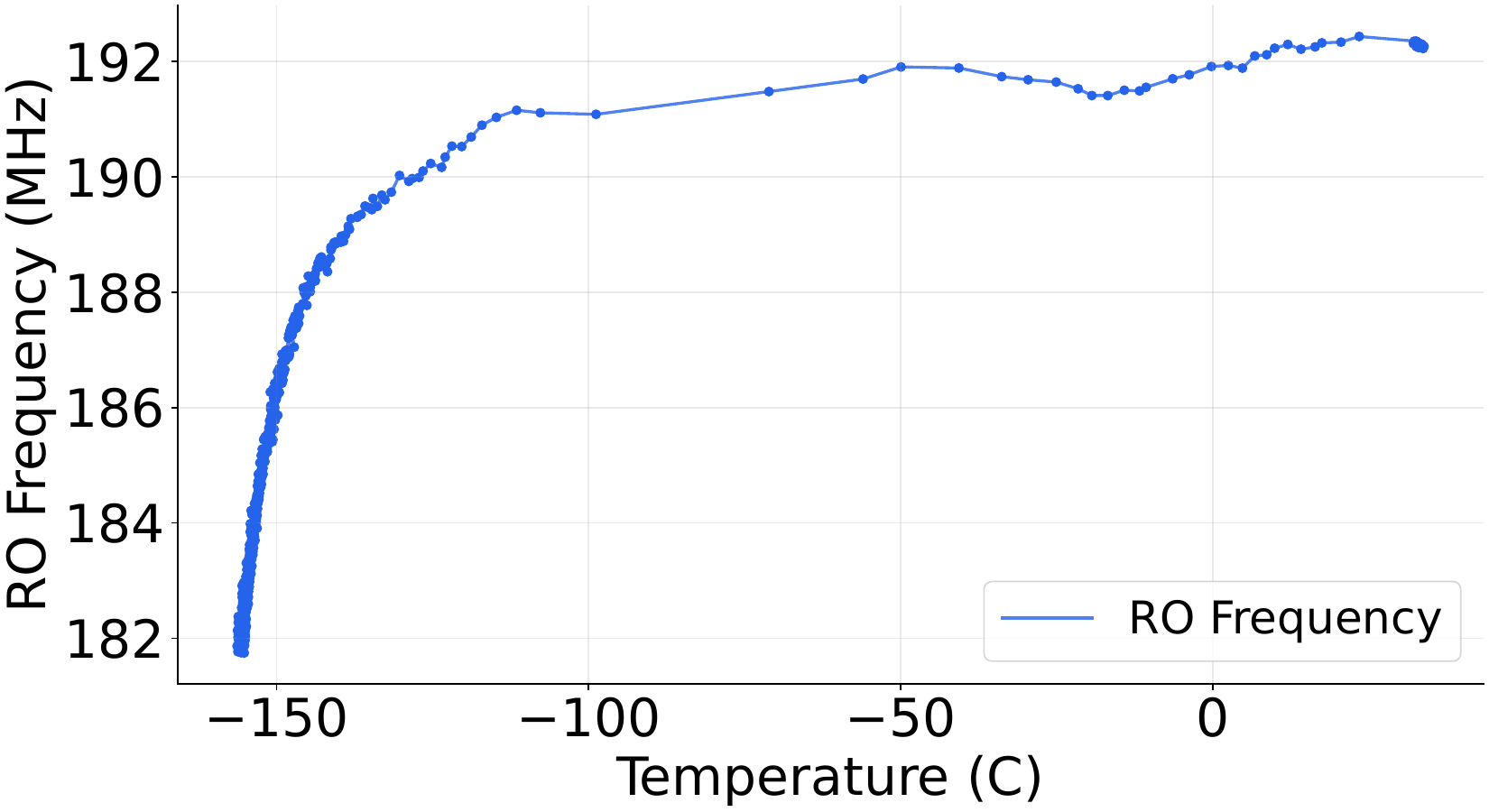}
        \caption{AMD Spartan-7}
        \label{fig:arty_ro_temp}
    \end{subfigure}

    \vspace{1em}

    \caption{RO frequency vs. temperature for the tested platforms under cryogenic conditions. For the AMD Spartan-7 FPGA and Zynq UltraScale+ MPSoC, the RO frequency decreases with temperature, indicating that the RO operates more slowly at lower temperatures. For the Microchip PolarFire FPGA, the RO frequency initially increases as the temperature decreases, reaches a maximum around -125$^\circ$C, and then decreases again at lower temperatures.}
    \label{fig:ro_temp_all}
\end{figure*}

\subsubsection{Clock Stability Monitoring}
To monitor the PLL behavior on the AMD Zynq UltraScale+ board, we used the MMCM IP and observed its lock signal. In this design, the MMCM input clock was generated from the PS clock and had a frequency of 33.33~MHz. The MMCM was configured to generate a 100~MHz output clock.

To determine how the input and output clocks behaved under cryogenic conditions, we implemented two 28-bit counters: one for the MMCM input clock and one for the MMCM output clock (See \cref{fig:cryogenic_test_environment_block_diagram}).
We used those counters to count the positive edges of the PLL input and output clocks. 
Comparing the four most significant bits (MSBs) of the counters enabled us to monitor and verify the PLL clock behavior at cryogenic temperatures.
By tracking the counter values, we monitored whether each clock continued running, slowed down, or stopped during the cryogenic experiment.

\subsubsection{Results}
For the AMD Zynq UltraScale+ device, we tested the chip in a cryogenic environment. Our observation was that at around -100$^\circ$C, the internal clock coming from the PS stopped working. As shown in \cref{fig:ultrascale_results}, the counter that was counting the clock coming from the PS, which was also used as the MMCM input clock, stopped counting. Similarly, the counter that counted the MMCM output clock edges also stopped. 
The 4 MSB bits of both counters show that they remained stuck at the last reported value. 
Also, the lock signal failed. These failures occur
due to \failanalog and perhaps \failcaps for the VCCAUX power rail for PS.

When the internal clock stopped, the entire design, including the MMCM output clock, froze, and the SYSMON temperature and voltage readings remained at their last reported values. Therefore, even though the chip temperature was further reduced, the reported temperature and voltage values could not reflect that change.
Also, although the I/O stopped working at a cryogenic temperature, it started working again after we warmed the chip back up. However, the internal clock generated by the PS remained stopped.


We also implemented an RO to observe the delay behavior in the 16 nm technology. As shown in \cref{fig:kv260_ro_temp}, the RO frequency decreases as the temperature drops. This indicates that at lower temperatures, the RO becomes slower due to the ITD effect.
Because the XADC stopped working at around -100$^\circ$C when using the internal clock, we performed the RO experiments using a stable external clock source to reliably communicate with the chip to read back the RO counter and record the temperature. This allowed us to measure the RO frequency as a function of temperature.

\subsection{Microchip PolarFire Results}
\subsubsection{Temperature and Voltage Sensor (TVS) Configuration}
Each PolarFire device includes a built-in temperature and voltage sensor (TVS), which can be used to monitor the internal die temperature and the main device supply rails~\cite{microchip_tvs_2022}. The TVS provides this information to the FPGA fabric in digital form. It is implemented with a 4-channel ADC, with each channel corresponding to a specific measurement. Channel 0 monitors the VDD (1 V) supply, Channel 1 monitors the VDDAUX (1.8 V) supply, Channel 2 monitors the VDDI (2.5 V) supply, and Channel 3 monitors the die temperature. The TVS outputs a 16-bit value. These raw digital values are then converted to standard voltage and temperature values for analysis.
In our design, we monitored the temperature and the VDD and VDDAUX voltage rails.




\subsubsection{Clock Resources}
For the Microchip PolarFire board, we used the 2~MHz clock generated by the internal PolarFire RC oscillator. This internal clock served as the clock source for both the TVS module and the Clock Conditioning Circuitry (CCC) input.

\subsubsection{Clock Stability Monitor}
To monitor the clock behavior on the PolarFire device, we used the CCC and tracked its lock signal. In this design, the CCC input was driven by the 2~MHz clock generated from the internal RC oscillator, and the CCC was configured to generate a 2~MHz output clock.
To evaluate the behavior of both the CCC input and output clocks under cryogenic conditions, we implemented two 28-bit counters: one for the CCC input clock, generated by the RC oscillator, and one for the CCC output clock, as shown in \cref{fig:cryogenic_test_environment_block_diagram}. These counters counted the positive edges of the corresponding clocks. By comparing the four most significant bits (MSBs) of the counters, we monitored the CCC clock behavior and verified whether the clocks continued running, slowed down, or stopped during the cryogenic experiment.
\begin{table*}[t]
\centering
\caption{Comparison of tested platforms under cryogenic conditions.}
\label{tab:platform-comparison}
\small
\begin{tabular}{>{\arraybackslash}p{3.0cm}>{\arraybackslash}p{4.0cm}>{\centering\arraybackslash}p{2.0cm}>{\arraybackslash}p{2.2cm}>{\arraybackslash}p{1.5cm}>{\arraybackslash}p{0.7cm}}
\toprule
&  & \textbf{Clock failure} & \textbf{Inferred} & \textbf{Sensors} &  \\
\textbf{Platform} & \textbf{Technology / package} & \textbf{temperature} & \textbf{failure modes} & \textbf{disabled?} & \textbf{ITD?} \\
\midrule
AMD UltraScale+ & 16\,nm, SRAM, flip-chip & $-100$\,\textdegree C & \circled{3}, \circled{4} & Yes & Yes \\
Microchip PolarFire & 28\,nm, Flash, BGA & $-150$\,\textdegree C & \circled{1}, \circled{2}, \circled{4} & Yes$^{*}$ & No \\
AMD Spartan-7 & 28\,nm, SRAM, BGA & $-176$\,\textdegree C & \circled{1}, \circled{3}, \circled{4} & Yes & Yes \\
\bottomrule
\multicolumn{6}{l}{$^{*}$ TVS did not recover on warm-up; reported invalid values ({\tt 0xFFFF}).} \\
\end{tabular}
\end{table*}

\subsubsection{Results}
For the Microchip PolarFire, we tested the chip in a cryogenic environment. Our observation was that at around -150$^\circ$C, the chip started to behave differently.
As shown in \cref{fig:PolarFire_results}, when the device reached cryogenic temperature, the TVS module started reporting invalid values, between approximately 40~s and 70~s. During this period, other parts of the design, including the clocks, RO, and monitoring logic, were still operating. This  is due to \failure{1}{VRM} and \failcaps for the VDDAUX power rail under cryogenic conditions.
Shortly after this interval, the device stopped responding. At this point, communication between the host PC and the chip was lost. 
Our external VDD measurement reported a voltage below 0.9 V, which, according to~\cite{mitard2025chypnosis}, indicates that the PolarFire device will enter a brownout condition and hibernate.
In this condition, the transistors on the digital fabric fail to switch, and hence the clock propagation halts.
The instability of VDD indicates that this Microchip PolarFire family, unlike the AMD Ultrascale+ family, does not exhibit ITD; therefore, cold temperatures cause faster transistor switching and current surges, leading to a significant voltage drop on the VDD power rail. 
Based on our observation, the VRM was unable to keep up with the higher current draw, and thus, the voltage could not be compensated. 
As a result, we can attribute this failure to \failure{1}{VRM} and \faildigitals.

The device could be awakened when it was warmed back up at around 410~s (see \cref{fig:PolarFire_results}), after which the chip started operating again, and communication was restored. Also, the counters for both the oscillator clock and the CCC output clock resumed operation. 
Since we monitored the 4 MSBs of the counters, we observed that both counters continued from the same values they had before the device stopped responding. This suggests that the chip was indeed in hibernation mode from around 70~s to 410~s, rather than being fully reset or reprogrammed.

It is noteworthy that when the device was warmed up again, the TVS sensor reported invalid values for both temperature and voltage rails.
The TVS voltage output was 0xFFFF. Based on our decoding function, this raw value is interpreted as $-4.095875$V. However, this value is not physically meaningful for voltage rails, since the supply voltage cannot be negative. Therefore, we treat this as an invalid TVS reading.
This indicates that the TVS sensor did not recover properly after being exposed to cryogenic temperatures.
Also, it is notable that the TVS high-temperature alarm was not triggered, even when the TVS sensor reported invalid high-temperature values. This suggests that the invalid TVS readings did not necessarily activate the corresponding tamper alarm flags.

Finally, the RO monitoring results on the Microchip PolarFire device show a non-monotonic behavior under cryogenic conditions (see \cref{fig:pf_ro_temp}). As the temperature decreased, the RO frequency initially increased, indicating that the RO became faster, confirming that this device is not exhibiting ITD at moderately cold temperatures. 
This trend continued until approximately -125$^\circ$C. Below this temperature, the RO frequency began to decrease, indicating that the RO slowed again at lower temperatures, indicating the appearance of the ITD effect at extremely cold temperatures.
Note that we report RO values only for time periods when the chip was not in hibernation mode. 
During hibernation, both the clock and the RO stopped operating, so the RO counter also stopped counting.

\subsection{AMD Spartan-7 Results}

\subsubsection{XADC Configuration}
The XADC includes a dual 12-bit, 1 Mega-sample-per-second (MSPS) analog-to-digital converter (ADC) and on-chip sensors~\cite{amd_xadc_ug480}. The ADCs provide a high-precision analog interface for various analog inputs. In addition to external analog inputs, the XADC can measure internal FPGA parameters, such as the device supply voltages and die temperature. 
The XADC data can also be accessed through the JTAG TAP interface, either before or after FPGA configuration~\cite{amd_xadc_ug480}. 
In our design, we instantiated the XADC module in Verilog and used it to read three internal measurements: die temperature, VCCINT, and VCCAUX. VCCINT is the FPGA core supply voltage, which is nominally 1.0 V, while VCCAUX is the auxiliary supply voltage, which is nominally 1.8 V.
The XADC has an input clock that drives its internal operation. 

\subsubsection{Clock Configurations}
For the AMD Spartan-7 board, we used the on-board clock generated by the board oscillator. This clock provides a 12 MHz input to the FPGA and serves as the main clock source in our cryogenic experiments.
Thus, we used this on-board clock as the input clock for both the XADC and the MMCM. (See \cref{fig:cryogenic_test_environment_block_diagram})

\subsubsection{Clock Stability Monitoring}
For the AMD Spartan-7 board, we used the MMCM IP as the clock-monitoring block and tracked its lock signal during the cryogenic experiment. The MMCM was driven by the on-board 12~MHz clock and configured to generate a 6~MHz output clock.

To study the behavior of both sides of the MMCM, we added two 28-bit counters, as shown in \cref{fig:cryogenic_test_environment_block_diagram}. One counter was driven by the MMCM input clock, while the other was driven by the MMCM output clock. These counters counted the positive clock edges, allowing us to monitor whether each clock continued operating, slowed down, or stopped as the device temperature decreased.
Instead of reading the full counter values, we recorded the four most significant bits (MSBs) of each counter. This provided a compact way to track clock activity and verify MMCM behavior under cryogenic conditions.

\begin{figure}
    \centering
    \includegraphics[width=0.80\linewidth]{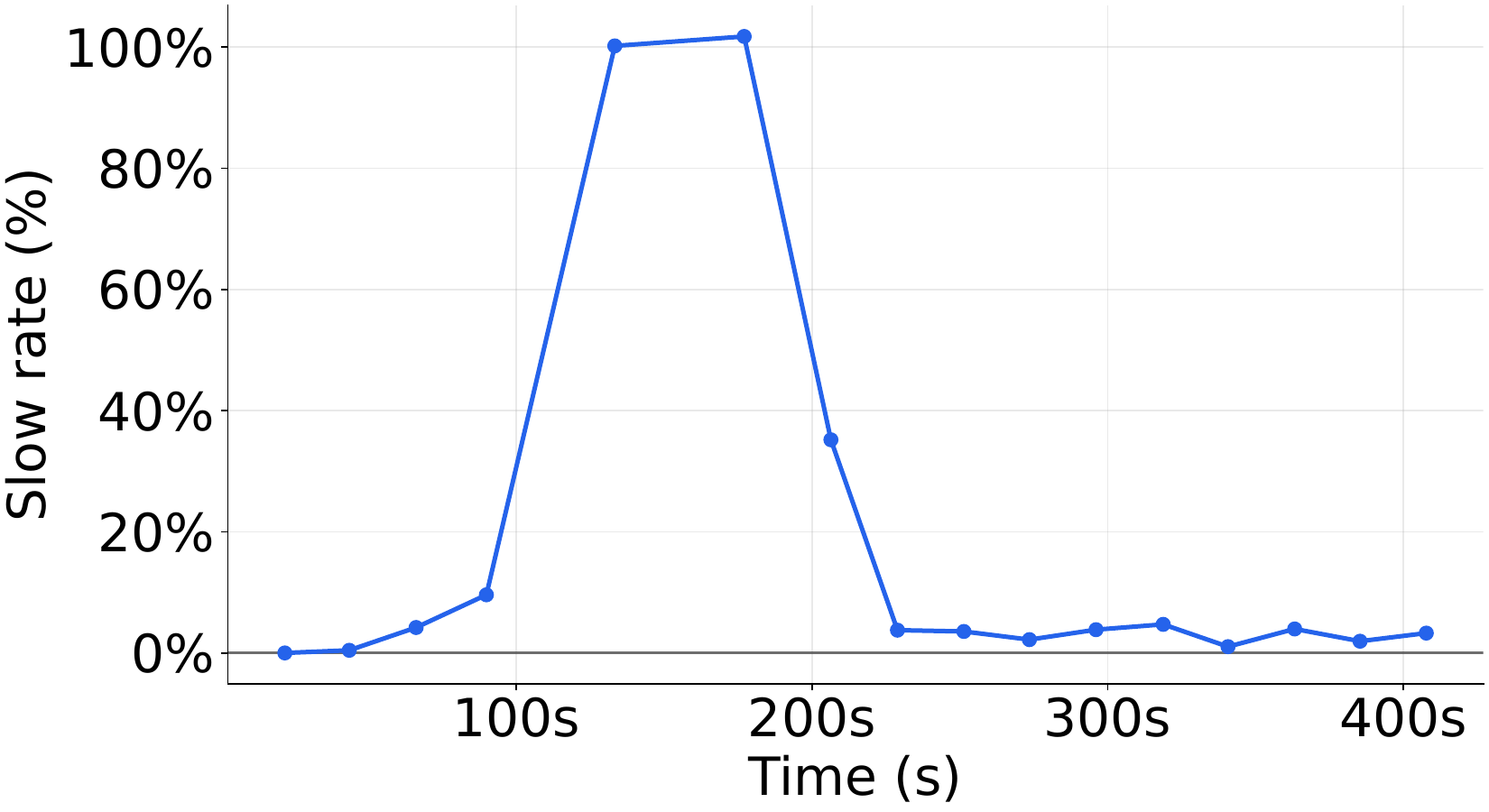}
    \caption{AMD Spartan-7 PLL input clock slow rate over time, calculated from changes in the input clock counter cycle period relative to the initial cycle period.}
    \label{fig:spartan7_slowrate}
\end{figure}

\begin{figure*}[t]
    \centering
    \includegraphics[width=0.9\linewidth]{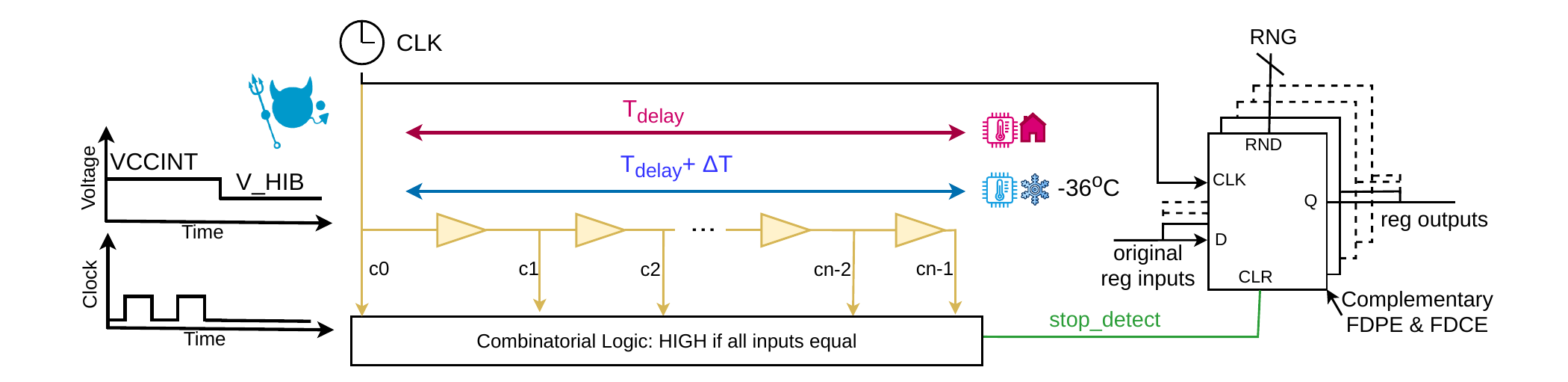}
    \caption{
Combined attack scenario against the state-of-the-art clock sensor proposed in~\cite {mitard2025chypnosis,dumitru2025borrowed,BorrowedTime_github}: The attacker first lowers the device temperature to -36$^\circ$C using Chypothermia, and then applies a fast voltage drop using Chypnosis~\cite{mitard2025chypnosis} to stop the clock. Lowering the temperature increases the delay chain due to Digital Malfunction caused by ITD, as shown in yellow. The additional delay introduced by cooling the system is represented by $\Delta T$. This extra delay extends the detection window and gives the attacker more time to stop the clock before the sensor asserts the \texttt{stop\_detect} signal, shown in green.
    }
    \label{fig:sensor}
\end{figure*}

\subsubsection{Results}
After placing the chip in a cryogenic environment, we observed that the PLL output clock stopped working when the chip temperature decreased to around -176$^\circ$C. As shown in \cref{fig:spartan7_results}, the PLL output clock counter stopped at a specific value, so its 4 MSB bits remained constant. Also, the lock signal failed to detect this abnormality.
In this case, our external monitoring VCCAUX starts showing first unreliable fluctuations and then a significant increase in the voltage level.
This observation is consistent with the VCCAUX surge reported in~\cite{homulle2019cryogenic}.
Such a voltage surge is due to possible failures in the analog biasing of PLLs~\cite{homulle2019cryogenic}, as well as unreliability in VRMs and capacitors under cryogenic conditions.
Hence, the PLL failure is caused by \failure{1}{VRM}, \failure{3}{Analog}  and \failcaps.

After pouring liquid nitrogen onto the chip's package, we observed that the PLL input clock started to slow down. \cref{fig:spartan7_slowrate} shows the slowdown rate of the PLL input clock over time, which was calculated from the change in the input-clock counter period relative to its initial period. Also, in \cref{fig:spartan7_results}, we show only the regions where the clock slowdown exceeded 10\%.
We also observed that this slowdown was temporary. The input clock slowed for a short period, then returned to its previous behavior. 
In this experiment, the clock source was the on-board oscillator, which was not directly exposed to liquid nitrogen. 

For the XADC sensors, we observed that after briefly cooling the chip, the sensors became deactivated and reactivated while the chip remained in the cryogenic environment. 
\cref{fig:spartan7_results} shows the sensor results for the AMD Spartan-7 device.
We also tracked the RO behavior in the AMD Spartan-7 device. As shown in \cref{fig:arty_ro_temp}, the RO starts to slow down as the temperature decreases.

\subsection{Peripheral Interface Reliability Under Cryogenic Conditions}
Based on our observations, the peripheral interfaces were affected when the devices were cooled to cryogenic temperatures. For the AMD Spartan-7 and Zynq UltraScale+ devices, the JTAG interface failed at very low temperatures, making it impossible to reprogram the chips. Specifically, we could not reprogram the AMD Spartan-7 at around -176$^\circ$C or the AMD Zynq UltraScale+ at around -168$^\circ$C. For the Microchip PolarFire device, the chip entered hibernation mode at cryogenic temperatures, disabling the device and rendering the interfaces unreachable.

The I/O pins on the AMD devices continued to operate during the cryogenic experiments. However, for the AMD Zynq UltraScale+ device, the interface became disabled after approximately 30 seconds of operation at cryogenic temperature. For the Microchip PolarFire device, once the chip entered hibernation mode, the I/O pins and other interfaces were inaccessible at that temperature.
After warming the devices back up, the JTAG and I/O interfaces started working again for all tested chips. This shows that the failures were temperature-dependent and reversible after returning the devices to a normal operating temperature.

\subsection{Results Conclusion}

Based on the results presented, we observe that both the clock source/synthesis circuits and the PLL-based clock sensors on all three chips are disabled at cryogenic temperatures, albeit for different reasons (see \cref{tab:platform-comparison}). 
Meanwhile, the continuous operation of the RO and the error-free states of the counters indicate that the FPGA's digital fabric remains operational or, if hibernated, retains its data intact for a later recovery using a static side-channel attack. 
This shows that the configuration of the FPGAs also remained intact.
Note that for chips, where the clock continues to operate upon warm-up, the adversary can manually place or keep the chip in hibernation at cryogenic temperatures by undervolting it~\cite{mitard2025chypnosis}, preventing the clock signal from being reactivated at warmer temperatures.
While we exposed only the chip's package and its surroundings to LN$_2$, heat from other parts of the PCB is also being extracted through the PCB's ground plane; therefore, the temperatures of the capacitors and VRMs are also lowered to cryogenic levels.
This is why we also observe capacitor and VRM malfunctions.
Finally, we observe that the temperature drop in flip-chip BGA packages was slower than that of traditional BGA packages.
This could be because the front-side cooling of the chip can more quickly extract heat from the transistors via the metal layers than the backside silicon.

\section{Combined Attack Results}

Reducing the temperature is a slow process, making the attack detectable if temperature sensors have been configured.
One can still disable the clock using Chypnosis~\cite{mitard2025chypnosis} attacks.
But such attacks can also be mitigated by asynchronous sensors, as proposed in~\cite{mitard2025chypnosis,AMD-SB-8018,PSIRT-118}.
In this section, we show that, to bypass the clock sensor, one can first cool the chip to moderate temperatures within the allowed thresholds, thereby slowing the sensor's sensing module (see \cref{fig:sensor}) due to \faildigitals caused by the ITD effect, and then mount Chypnosis to bypass the countermeasure.

\subsection{OpenTitan Implementation}
\begin{figure}[t]
    \centering
    \includegraphics[width=0.85\linewidth]{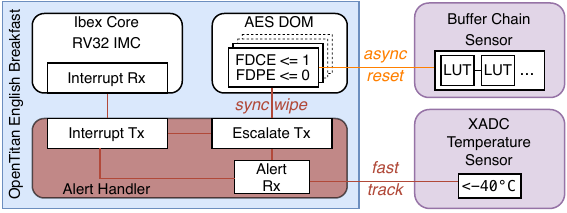}
    \caption{Architecture of trimmed OpenTitan English Breakfast platform, equipped with an alert handler for synchronously wiping and a buffer chain sensor for asynchronously resetting registers.
    }
    \label{fig:opentitan}
\end{figure}

To demonstrate the effectiveness of our combined attack, we target the OpenTitan English Breakfast version~\cite{opentitan_englishbreakfast}, a root-of-trust platform that integrates a 2-share Domain-Oriented Masking (DOM)~\cite{gross2016domain} AES implementation designed to defend against side-channel attacks.
The objective is to mount a combined attack to bypass detection and recover the secret key stored in the registers of the secure AES module.
For the experimental platform, we equipped OpenTitan with an alert handler~\cite{opentitan} and a buffer chain sensor from~\cite{mitard2025chypnosis,BorrowedTime_github} with a complementary register design, as shown in \cref{fig:sensor}, to erase critical register values upon detecting the Chypnosis attack.
The architecture of the platform is shown in \cref{fig:opentitan}.
For our experiment, we trimmed modules without degrading any security aspect, such as removing the 192-bit encryption mode within the AES module, to implement OpenTitan on our experimental FPGA boards due to limited hardware resources.

The alert handler is responsible for detecting abnormal temperature conditions through the connection to the XADC module.
We configure the alert handler in fast-track response mode, so that when the measured temperature drops below -40$^{\circ}$C, the response is triggered in only four clock cycles.
Since temperature reduction is a relatively slow process, this configuration allows the alert handler to erase sensitive registers synchronously before the attack can take effect.
In our design, the buffer chain-based sensor detects rapid voltage reductions and can erase sensitive data by incorporating an asynchronous reset mechanism.

\subsection{Attack Results}

During the cooling experiments, we observed that as the temperature decreased, the RO frequency in the AMD Spartan-7 device also decreased. This means that the design becomes slower at lower temperatures.
This behavior can help an attacker bypass delay-based security countermeasures. When the circuit slows down, the sensor's timing margin changes, which can make the countermeasure less sensitive to fast voltage drops (Chypnosis attack).

To evaluate this effect, we performed the combined Chypothermia and Chypnosis attack on the OpenTitan implementation on the AMD Spartan-7 board.
We set the XADC temperature threshold to -40$^\circ$C. So, if the temperature goes below -40$^\circ$C, the XADC temperature sensor should also detect it.
In addition, we used the delay-based sensor from~\cite {mitard2025chypnosis} to detect clock halts caused by voltage drop.
At room temperature, we observed that when we rapidly dropped the voltage, the sensor detected the attack and wiped the AES key. 
However, by placing the chip inside the thermal chamber and setting the chamber temperature to -36$^\circ$C, we then repeated the same voltage-drop attack. In this case, the detection did not occur, and the AES key remained intact in the key registers, which were read out after the chip was awakened and the clock was re-enabled.
This shows that reducing the temperature can slow down the countermeasure, making it vulnerable.

\subsection{Results Conclusion}
We showed that reducing the temperature can increase the delay of delay-based sensors, enabling an attacker to stop the clock without triggering the detection mechanism and preserving the secret data on the chip.
Note that this applies to devices that exhibit ITD behavior, which includes most modern chips with technologies smaller than 28~nm.
Otherwise, the temperature-delay behavior can be inverted, and cold temperatures make the sensor faster, as is the case for Microchip devices, see \cref{fig:ro_temp_all}. 
In such devices, an adversary may need to actually heat the chip rather than cool it to increase the delay and execute a similar attack.
Also, it is important to note that, according to the literature~\cite{glamovcanin2023temperature,mehta2024bake}, AMD 7-Series devices exhibit both ITD and non-ITD behaviors, depending on how the sensor is placed and routed on the chip.
In this work, our sensor location on the FPGA followed the ITD pattern, and therefore, we could attack it.
\section{Countermeasures}\label{sec:counter}

\subsection{Self-heating Chip}


\begin{figure}[t]
  \centering
  \includegraphics[width=0.8\linewidth]{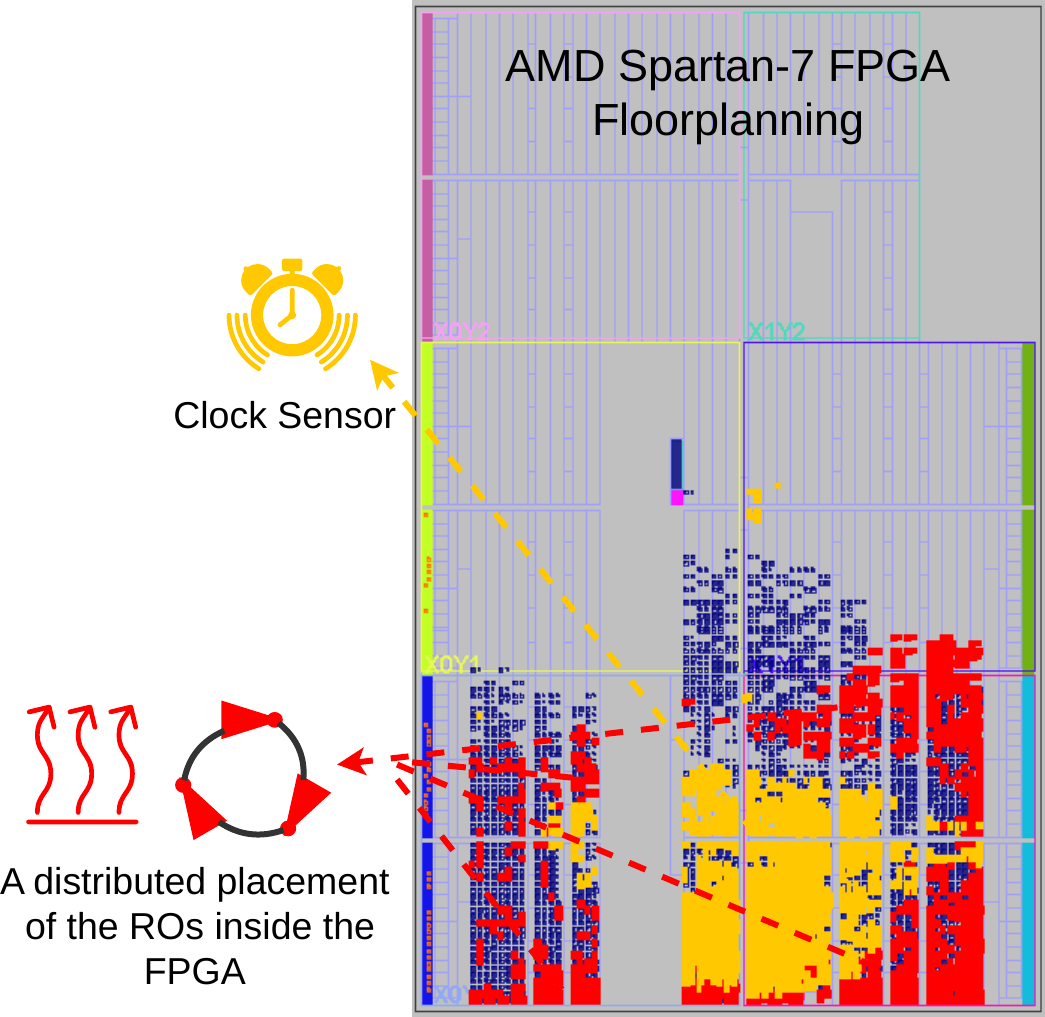}
  \caption{Implementation of the self-heating countermeasure on an AMD Spartan-7 FPGA. The delay-based clock-stop sensor (yellow) is surrounded by a distributed deployment of $1500$ three-stage ROs (red), 
  keeping the sensor within its nominal operating-temperature range.}
  \label{fig:RO_countermeasure_figure}
\end{figure}

Our combined attack succeeds because low temperatures affect the signal propagation delay of the sensor's sensing module, i.e., the buffer chain.
A natural countermeasure would be to actively heat the sensor from within to keep it within its nominal delay regime. 
Ring oscillators (ROs) are well-suited to this task, as they are built entirely from standard CMOS logic, are compatible with FPGAs, and can be placed almost anywhere on the die. 
They are among the most efficient on-chip heat generators reported, raising the die temperature by more than 130$^\circ$C while occupying only about $21\%$ of the slices on FPGAs~\cite{happe2012eight}. 
Their self-heating remains substantial even at deep-cryogenic conditions, where a CMOS ring oscillator has been measured to self-heat by more than $150$~K (i.e., a rise exceeding 150$^\circ$C) at a $4.2$~K (-268.95$^\circ$C) bath depending on its power dissipation~\cite{hamlet2011modeling}. 
More broadly, the principle of pairing such logic-based heaters with an on-chip temperature sensor and a controller to hold a temperature-sensitive circuit at a predefined operating point is well established~\cite{hars2020onchip}. 
We therefore surround the temperature-sensitive clock-stop sensor with a farm of ROs that, once activated, generate sufficient localized heat to keep the sensor responsive, thereby mitigating cooling-based attacks.

Following the approach of~\cite{hars2020onchip}, the RO farm is not permanently enabled; instead, it is activated automatically once the on-die temperature reported by the XADC sensor falls below a predefined threshold.
In this condition, it heats the surrounding logic to pull the sensor back into its safe operating temperature range. 
We set the threshold to -29$^\circ$C, corresponding to the die temperature reported by the XADC when the thermal chamber is set to -36$^\circ$C. At this operating point, the die remains approximately 7$^\circ$C warmer than the ambient, staying above the -40$^\circ$C anti-tamper threshold and avoiding a temperature alarm, while still being cold enough to slow the delay chain and disable the clock-stop sensor.

\cref{fig:RO_countermeasure_figure} shows our implementation on a modified AMD Spartan-7 board, where a distributed farm of three-stage ring oscillators is placed across the fabric around the delay-based clock-stop sensor that produces \texttt{stop\_detect}, so that the generated heat flows to the sensor placement on the chip.
To size the heater, we held the chamber ambient at the same -36$^\circ$C used in the attack and swept the number of active ROs from $0$ to $2500$ while recording the resulting steady-state die temperature with the XADC. \cref{tab:ro_heating} reports these measurements, which rise with the number of active oscillators, from -29$^\circ$C with no ROs active to -18$^\circ$C at $2500$ ROs. 
We found that activating $1500$ three-stage ROs, which raises the die temperature to about -22.5$^\circ$C, is sufficient to keep the delay chain in its nominal regime; we therefore deploy $1500$ ROs in the final implementation of \cref{fig:RO_countermeasure_figure}. With the heater active, the sensor again asserts \texttt{stop\_detect} within its expected timing window, triggering the zeroization response and mitigating the combined Chypothermia-Chypnosis attack.

Deploying ROs as a countermeasure introduces both area and power overhead. In our implementation, each RO had three stages, so implementing 1500 ROs required approximately 4500 LUTs.
Naturally, increasing the number of ROs requires increasing LUT utilization.
We also measured the power overhead of the RO farm on the AMD Spartan-7 FPGA core supply rail. With the deployed $1500$ ROs active, the core consumes approximately $323$~mW, compared to roughly $30$~mW when the farm is disabled, i.e., roughly $0.2$~mW per oscillator. 
Since the ROs are enabled only below a certain temperature threshold, the countermeasure imposes no power overhead on the main design under nominal temperature ranges.

\begin{table}[t]
  \setlength{\belowcaptionskip}{2pt}
  \centering
  \small
  \caption{Measured steady-state on-die temperature (reported by the XADC sensor) of the AMD Spartan-7 versus the number of active three-stage ROs. The thermal chamber maintains the ambient at -36$^\circ$C throughout; with no ROs active, the die self-heats to -29$^\circ$C, which is above the operating point of the combined attack.}
  \label{tab:ro_heating}
  \small
  \begin{tabularx}{\columnwidth}{@{}>{\centering\arraybackslash}X>{\centering\arraybackslash}X>{\centering\arraybackslash}X@{}}
    \toprule
    \textbf{Active ROs} & \textbf{Chamber / ambient ($^\circ$C)} & \textbf{Die temperature ($^\circ$C)} \\
    \midrule
    0    & $-36$ & $-29.0$ \\
    1000 & $-36$ & $-25.1$ \\
    1500 & $-36$ & $-22.5$ \\
    2000 & $-36$ & $-20.0$ \\
    2500 & $-36$ & $-18.0$ \\
    \bottomrule
  \end{tabularx}
\end{table}

\subsection{Other Potential Countermeasures}

A complementary direction is to harden the off-chip components for cryogenic operation rather than only heating the sensor. Established cryogenic-electronics design shows the feasibility of reliable operation with carefully chosen custom parts, allowing an FPGA system to remain functional down to $4$\,K~\cite{homulle2018design,homulle2019cryogenic}. High-permittivity ceramic decoupling is a poor choice in cryogenic temperatures, since it loses most of its capacitance and gains series resistance. 
Therefore, the design should instead use NP0/COG (low-dielectric constant materials), PPS (polyphenylene sulfide film dielectric), or tantalum capacitors~\cite{homulle2018design,homulle2019cryogenic,teverovsky2006performance}.
On the other hand, since commercial voltage regulators stop working below roughly $90$\,K, the supply could be regulated by deploying improved voltage regulators~\cite{lewis2025implementation}, using internal ROs, or feeding it from a room-temperature power source~\cite{homulle2018design,homulle2019cryogenic}. 
Similarly, the clock can be generated by a cryogenic crystal oscillator or routed in from a source at room temperature.

Off-the-shelf military- and defense-grade chips could also be more reliable, as they are screened and qualified for much wider operating temperature ranges than commercial parts~\cite{xilinx7seriesdefense}. 
The advantage of these board-level and device-level choices is that they reuse mature, commercially available components and keep the platform's operation predictable across a wide range of temperatures.
However, their main drawback is the cost associated with cryo-grade and ruggedized parts. 
Note that while these choices could prevent failure of mixed-signal components on specific chips, they cannot be generalized to all chip families. For instance, it has been shown in~\cite{homulle2017performance} that even with cryo-grade components, the mixed-signal behavior of some FPGA families remains unreliable under cryogenic conditions.
Moreover, the effect of cooling on the digital circuits could still threaten the system's security, as discussed in our combined attack scenario.
Hence, the board-level solutions are most effective when combined with the active self-heating approach presented in the previous section.

Another way to counter Chypothermia and Chypnosis attacks is to block or detect unauthorized access to the chip’s package and its supply voltage using PCB-level secure enclosures~\cite{immler2018b,immler2019secure,staat2022anti} or tamper sensors~\cite{mosavirik2023impedanceverif}. Once tampering is detected, they can erase sensitive data before an attacker can carry out a static SCA attack.

\section{Discussion}\label{sec:discussion}

\subsection{Comparison with Data Remanence Attacks}
It may seem that our proposed attack resembles data remanence~\cite{oren2013effectiveness,anagnostopoulos2018low}, Cold Boot~\cite{halderman2009lest}, or Pentimento~\cite{drewes2024pentimento} attacks, where the adversary leverages charge retention or bias temperature instability in transistors to reconstruct data that used to reside in memory. Nevertheless, our attack is distinct in an important aspect.
In those prior attacks, it is typically assumed that the adversary can execute their own firmware or bitstream on the device after the sensitive data has been erased, and then exploit the analog properties of the memory (e.g., SRAM metastability, SRAM power-up patterns, or flip-flop propagation delays) to retrieve the original contents.
In contrast, our approach directly targets memory content that is retained under extremely cold conditions and assumes that the data can be extracted using static side-channel attacks, such as LLSI~\cite{mitard2025chypnosis,krachenfels2021real} or IA~\cite{mitard2025chypnosis,monfared2023leakyohm}, without requiring the adversary to take control of the chip by executing code and reading back data.

\subsection{Applicability to ASICs}
Since we performed our attacks on FPGA SoC platforms, questions may arise regarding their applicability to ASICs.
First, it is important to note that our cryogenic attack mainly targets mixed-signal components on chips, which are separate IPs from the digital FPGA fabric and are similar to those implemented on ASICs. 
Second, for the combined attack, we exploited the ITD effect, which is common to modern chips and not specific to FPGAs.
Hence, although an in-depth comparison of FPGAs and ASICs lies outside the scope of this work, our results provide strong evidence that the fundamental mechanisms exploited in our attack are not unique to FPGAs.
\section{Conclusion}\label{sec:conclusion}
In this paper, we presented Chypothermia, a cryogenic attack that exploits failures in on-chip mixed-signal components at cryogenic temperatures to disable clock sensors, voltage sensors, and clock-generation circuitry without requiring any electrical tampering of the target system. 
Moreover, we demonstrated that by combining Chypothermia with Chypnosis, clock halting can be achieved even at moderately low operating temperatures while evading thermal anomaly detection mechanisms. 
Our evaluation across multiple FPGA/SoC platforms showed that both soft-IP and hard-IP sensing implementations can be disabled while preserving the secret on the chip. 
We further demonstrated the practical impact of the attack on the OpenTitan root of trust, in which our combined attack successfully bypassed a state-of-the-art clock-monitoring architecture and prevented key zeroization. 
Our results again show that the security of chips' anti-tamper sensors remains underexplored; we cannot simply assume that conventional reliability sensors on the chip can withstand adversarial manipulation of environmental conditions.
Our results highlight the need for more resilient sensor designs and protection strategies that remain trustworthy under extreme environmental conditions.

\section{Responsible Disclosure} 
Following the discovery of the vulnerability, we responsibly disclosed it to AMD and Microchip on June 15, 2026, upon completing the initial version of the manuscript. We informed AMD and Microchip on the same date that we planned to keep the paper under embargo for 45 days. Microchip acknowledged the report on June 15, 2026. AMD acknowledged the report on June 18, 2026. All parties have remained in contact since the paper report was sent out. No further embargo was requested by either Microchip or AMD in follow-up communications.

\section*{Acknowledgments}
This effort was sponsored in part by NSF Grants CNS-2150123 and CNS-2338069 and in part by 
an ARC Discovery Project number DP210102670,  
the Deutsche Forschungsgemeinschaft (DFG, German Research Foundation) under Germany’s Excellence Strategy - EXC 2092 CASA - 390781972
and DFG project number 560392681.

\bibliographystyle{IEEEtranSN}
\bibliography{ref_fixed_compt}

\end{document}